\documentclass[journal=jctcce,manuscript=article]{achemso}
\usepackage{amssymb,amsmath,amsfonts,multicol,multirow,longtable,array,mathpazo}
\usepackage{lscape}
\usepackage[version=3]{mhchem} 
\usepackage{appendix}
\usepackage{soul} 
\usepackage[usenames]{color}
\usepackage{makecell}
\usepackage{enumerate}
\usepackage{xr}
\usepackage[T1]{fontenc} 
\usepackage{booktabs}
\usepackage{tikz}
\usetikzlibrary{shapes.geometric}
\usepackage{threeparttable}
\usepackage{graphicx}
\usepackage{subfigure}
\usepackage{colortbl}
\usepackage{framed}
\usepackage{verbatim}
\usepackage{amsbsy}
\usepackage{bm}
\usepackage{bbm}
\usepackage[normalem]{ulem}
\usepackage{float}
\usepackage{float}
\usepackage{algorithm}
\usepackage{algpseudocode}
\usepackage{subfigure}
\usepackage{simplewick} 
\usepackage{siunitx}

\usepackage{xr}

\newcounter{xscheme}

\newfloat{Algorithm}{htbp}{alg}
\floatname{Algorithm}{Algorithm}

\newcounter{exe}[figure]
\newcommand{\iexe}{\refstepcounter{exe}\the\value{exe}:}

\setkeys{acs}{maxauthors = 0} 

\author{Ning Zhang}\email{ningzhang1024@gmail.com}
\author{Wenjian Liu}\email{liuwj@sdu.edu.cn}
\affiliation{Qingdao Institute for Theoretical and Computational Sciences and Center for Optics Research and Engineering,
	Shandong University, Qingdao, Shandong 266237, China}

\title{cQED-iCIPT2: A Near-Exact Method for Polaritonic Chemistry}

\setkeys{acs}{articletitle=true}

\begin{document}
	
\begin{abstract}
Strong light-matter coupling in optical cavities provides a versatile platform for modulating chemical structure, reactivity, and spectroscopy,
and hence motivates the development of ab initio cavity quantum electrodynamics (cQED) methods
that can treat the electronic and photonic degrees of freedom on an equal footing.
We present such a method,  cQED-iCIPT2, by combining the near-exact iCIPT2 (iterative configuration interaction
with selection and second-order perturbation theory) with the cQED Hamiltonian in two ways,
cQED-PN-iCIPT2 and  cQED-CS-iCIPT2. The former works directly in the photon-number representation, whereas
the latter employs a coherent-state transformation that restores origin invariance of the cQED Hamiltonian and avoids artificially strong coupling in charged systems.
To efficiently handle the tensor-product structure of the electron-photon wavefunction, we introduce
a graded configuration space organized by photon numbers and decompose the key computational steps
(selection, diagonalization, and perturbation correction) into intra- and inter-subspace steps,
so as to maximize the reuse of the existing infrastructure in the MetaWave platform (J. Phys. Chem. A 2025, 129, 5170).
The selection step enables automatic determination of the optimal number of photons without \textit{a priori} truncation.
The efficacy of the methods is showcased dissociation of \ce{N2}, torsion of ethylene, proton transfer reactions
in malonaldehyde and aminopropenal, and low-lying excited states of polyacenes.
The results provide numerically accurate reference data and meanwhile reveal how the cavity can fine-tune reaction barriers, alter potential energy surfaces, and induce state crossings. As such, this work establishes a robust near-exact framework for polaritonic chemistry
in the presence of both strong electron correlation and strong light-matter coupling.
\end{abstract}

\maketitle

\clearpage
\newpage

\section{Introduction}


The manipulation of molecular properties via strong light–matter interaction in optical cavities
has emerged as a promising avenue in chemistry and materials science, and continues to be an active area of research.
By confining electromagnetic modes in optical cavities, coherent energy exchange between molecular excitations and vacuum fluctuations
can be enabled to yield hybrid light–matter states known as polaritons.
These polaritonic states exhibit modified absorption, emission, energy transport, and even chemical reactivity,
offering a non-invasive route to control ground- and excited-state potential energy surfaces\cite{QED_Review_Chemistry,QED_Review_Chemistry2,QED_Review_Current_Status,
QED_Review_MolecularPolaritonics,QED_Review_spectroscopies,cQEDExp2013,cQEDExp2015,cQEDExp2016,
cQEDExp2017,QED_BrominationReaction,QED_DAReaction,cQEDExp2021,cQEDExp2022,cQEDExp2023}.

The theoretical framework for a unified treatment of the electronic and photonic degrees of freedom
is \emph{ab initio} cavity quantum electrodynamics (cQED)\cite{QEDReview,QEDReview2,QED_REview,QED_REview2,QED_REview3,cQED_theo_review}.
The starting point of cQED is a mean-field description (QED-HF) of the light-matter interaction\cite{riso2022molecular,QED_MO_LargeSystem,QED_HF_AllStrength}.
On top of this, correlated density functional theory\cite{QEDFT,QEDFT_OEP1,QEDFT_OEP2}, time-dependent density functional theory\cite{QED_TDDFT_1,QED_TDDFT_2,QED_TDDFT_3,QED_TDDFT_4,cQED_4c_linear_response},
as well as single-reference (SR)\cite{QED_HF_Opt,QED_HF_response,QED_LF_MP2,QED_SC_MP2,QED_HF_CCSD,QED_CC,QED_CC_Large,QED_CC_2020,QED_RCC,cQED_CC_junjie,QED_DMRG}
and multi-reference (MR)\cite{QED_CASCI,QED_CASSCF_2025,QED_CASSCF_2025_2,QED_AFQMC,QED_DQMC,QED_QMC_THC} wave function-based methods have been developed in the last decade.
However, all the methods have to test manually the convergence with respect to the number of photons.
That is, there is still a lack of methods that can determine automatically the relevant photonic subspace
while handling efficiently strong electron correlation across large active spaces.
In this work, we propose such a method, cQED-iCIPT2, which is a combination of the near-exact iCIPT2 (iterative configuration interaction\cite{iCI} with selection and second-order perturbation theory)\cite{iCIPT2,iCIPT2New} with the Pauli–Fierz (PF) Hamiltonian\cite{QEDReview,QEDReview2}.
iCIPT2 belongs to the big family of sCIPT2 (selected configuration interaction plus second-order perturbation theory) methods,
which have proven remarkably successful in capturing both static and dynamic correlation for strongly correlated electrons.

The remainder of this paper is organized as follows. The basics of cQED is first reviewed briefly in
Sec.~\ref{sec:theory}, which is followed by the introduction and implementation of cQED-iCIPT2 in Sec. \ref{Sec-cQED-iCIPT2}.
Numerical results and concluding remarks are then provided in Sec. \ref{sec:pilot_app} and \ref{sec:conclusion}, respectively.

\section{Cavity QED}\label{sec:theory}
\subsection{Cavity QED Hamiltonian}
Under the Born–Oppenheimer and long-wavelength (dipole) approximations, the PF Hamiltonian\cite{QEDReview,QEDReview2} for a molecular system coupled to a single photonic mode
takes the following form (in the length gauge and atomic units)
\begin{align}
\hat{H}_{\mathrm{PF}} &= \hat{H}_{\text{e}} + \hat{H}_{\text{cav}} + \hat{H}_{\text{int}} + \hat{H}_{\text{DSE}},\label{eqn:PF_Hmat}\\
\hat{H}_{\text{cav}}&=\omega \hat{b}^{\dagger} \hat{b},\\
\hat{H}_{\text{int}}&=-\sqrt{\frac{\omega}{2}} \hat{d}\left(\hat{b}^{\dagger}+\hat{b}\right),
\quad \hat{d}=\bm{\lambda} \cdot \hat{\boldsymbol{\mu}}=\hat{d}_{\mathrm{e}}+d_{\mathrm{n}},\label{Hint}\\
\hat{H}_{\text{DSE}}&=\frac{1}{2}\hat{d}^2,\label{DSEdef}
\end{align}
where $\hat{H}_{\mathrm{e}}$ is the Born–Oppenheimer many-electron Schr\"odinger-Coulomb Hamiltonian, $\hat{H}_{\text{cav}}$
describes the bare cavity mode as a harmonic oscillator of fundamental frequency $\omega$ (with
$\hat{b}^{\dagger}$ and $\hat{b}$ being the creation and annihilation operators of the cavity photonic mode, respectively),
$\hat{H}_{\text{int}}$ describes the bilinear coupling between the light and matter,
whereas $\hat{H}_{\text{DSE}}$ represents the dipole self-energy (DSE), without which
the Hamiltonian \eqref{eqn:PF_Hmat} would not be bounded from below and the combined light–matter system would be unstable~\cite{QED_DSE1,QED_DSE2}.
The molecular dipole operator $\hat{\boldsymbol{\mu}}$ contains both
electronic ($\hat{\boldsymbol{\mu}}_{\mathrm{e}}$) and nuclear ($\boldsymbol{\mu}_{\mathrm{n}}$) contributions
(with the latter being constant for a given geometry).
The coupling vector can be written as $\bm{\lambda}=\sqrt{\frac{\hbar}{\varepsilon_0 V}} \hat{\mathbf{e}}$
in terms of the cavity volume $V$ and the polarization unit vector $\hat{\mathbf{e}}$.

The DSE term \eqref{DSEdef} can be expressed in two ways.
In the first, the squared operator is taken as first-quantized and then second-quantized (``second-quantized after product''):
\begin{align}
\hat{H}_{\text{DSE}}&=\frac{1}{2} \sum_{i\neq j}^{N_{\mathrm{e}}} (\boldsymbol{\lambda} \cdot \hat{\boldsymbol{\mu}}(i))(\boldsymbol{\lambda} \cdot \hat{\boldsymbol{\mu}}(j)) + \frac{1}{2}\sum_{i=1}^{N_{\mathrm{e}}} (\boldsymbol{\lambda} \cdot \hat{\boldsymbol{\mu}}(i))^2\nonumber \\
	&=\frac{1}{2}\sum_{pqrs} \bar{d}_{pq} \bar{d}_{rs}  e_{pq,rs}
	+\frac{1}{2}\sum_{pq} \bar{q}_{pq} E_{pq}, \label{eqn:dse_sp_form}\\
\bar{d}_{pq}&=\boldsymbol{\lambda} \cdot\langle \phi_p| \hat{\boldsymbol{\mu}}|\phi_q\rangle,\\
\bar{q}_{pq}&=\boldsymbol{\lambda} \cdot\langle\phi_p| \hat{\boldsymbol{q}}|\phi_q\rangle \cdot \boldsymbol{\lambda},\quad \hat{q}_{xy}=\hat{\mu}_x\hat{\mu}_y,
\end{align}
where $\hat{\boldsymbol{\mu}}(i)$ is the dipole operator of electron $i$,
and $\bar{d}_{pq}$ and $\bar{q}_{pq}$ are the (modified) dipole and quadrupole integrals, respectively.
This form does not assume the completeness of the electronic one-particle basis set.
In the second approach, the dipole operator is first second-quantized and then squared (``product after second-quantization''):
\begin{align}
\hat{H}_{\text{DSE}}&=\frac{1}{2}
		\left(\sum_{pq} \bar{d}_{pq} E_{pq}\right)
		\left(\sum_{rs} \bar{d}_{rs} E_{rs}\right)\nonumber \\
		& = \frac{1}{2}\sum_{pqrs} \bar{d}_{pq} \bar{d}_{rs} e_{pq,rs} + \frac{1}{2} \sum_{pq}\bar{q}_{pq}  E_{pq},\quad
\bar{q}_{pq}=\sum_r \bar{d}_{p r}\bar{d}_{r q},\label{eqn:dse_ps_form}
	\end{align}
which manifestly depends on the quality of the electronic one-particle basis set.
Nevertheless, the difference between the two forms of DSE 
occurs mainly to total energies but is negligibly small for energy differences
(see Sec. S1 in the Supporting Information). Therefore, only the former is employed in the present calculations.

To facilitate the subsequent discussion, we rewrite the PF Hamiltonian for polaritons in a generic form,
\begin{align}
\hat{H} & = \hat{h}_{\text{e}} + \hat{V}_{\text{ee}} + \hat{h}_{\text{p}} + \hat{h}_{\text{e-p}} \label{eqn:generic_elec_boson_h}\\
		&=\sum_{p q } h_{p q} E_{pq} +\frac{1}{2} \sum_{p q r s} V_{p q r s} e_{pq,rs}
		+\omega b^{\dagger} b
		+\sum_{p q} g_{p q} E_{pq} \left(b^{\dagger}+b\right),\label{eqn:generic_elec_boson_h-int}
\end{align}
where
\begin{align}
		E_{p q} & =\sum_\sigma a_{p \sigma}^{\dagger} a_{q \sigma}=E_{q p}^{\dagger} \\
		e_{pq, rs} & =\sum_{\sigma, \tau} a_{p \sigma}^{\dagger} a_{r \tau}^{\dagger} a_{s \tau} a_{q\sigma}=E_{pq} E_{rs}-\delta_{q r} E_{p s}=\left\{E_{p q} E_{r s}\right\} \nonumber\\
		& =\left\{E_{r s} E_{p q}\right\}=e_{rs, pq}=e_{qp, sr}^{\dagger},\\
g_{p q}&=-\sqrt{\frac{\omega}{2}} |\lambda|(\mathbf{e} \cdot \boldsymbol{\mu})_{p q},\\
\bm{\mu}_{pq} &= \bigl\langle \phi_p \big| \bigl( \mathbf{r} - \frac{\sum_I Z_I \mathbf{R}_I}{N_{\text{e}}} \bigr) \big| \phi_q \bigr\rangle.
	\label{eqn:dipole_int}
\end{align}
Here, the fermionic operator $a_{p \sigma}^{\dagger}$ creates an electron with spin $\sigma$ at orbital $p$,
the bosonic operator $b^{\dagger}$ creates a boson (here a photon, but the form is general),
$h_{p q}$ and $V_{p q r s}$ denote the respective one- and two-electron integrals,
$Z_I$, $\mathbf{R}_I$ and $N_{\text{e}}$ are the respective nuclear charge, nuclear position and number of electrons.
The generic Hamiltonian \eqref{eqn:generic_elec_boson_h} corresponds to
$\hat{H}_{\mathrm{PF}}$ \eqref{eqn:PF_Hmat} by setting 
 $\hat{h}_{\text{p}}=\hat{H}_{\text{cav}} $ and $\hat{h}_{\text{e}} + \hat{V}_{\text{ee}}=\hat{H}_{\mathrm{e}}+\hat{H}_{\mathrm{DSE}}$.
 The latter can be implemented by absorbing the DSE contribution in Eq. \eqref{eqn:dse_sp_form} or Eq. \eqref{eqn:dse_ps_form} into the one- and two-electron integrals in Eq.~\eqref{eqn:generic_elec_boson_h-int} as follows
\begin{equation}
	\begin{aligned}
		h_{pq} & \gets h_{pq} + \frac{1}{2}\bar{q}_{pq},  \quad
		V_{pq,rs}  \gets V_{pq,rs} + \bar{d}_{pq} \bar{d}_{rs}.
	\end{aligned}
\end{equation}

\subsection{Polaritonic Wavefunctions}

The simplest wavefunction for a polariton is the direct
product of a single Slater determinant \(|\Phi_{\mathrm{e}}\rangle\) of $N_{\mathrm{e}}$ electrons and the photon vacuum $|0_{\mathrm{p}}\rangle$.
However, such description misses all electron–photon couplings, which can only be resolved by taking a
linear combination of PN states for the photonic state $|\Phi_{\mathrm{p}}\rangle$, viz.
\begin{equation}
	|\Psi_0\rangle\equiv\left|\Phi_{\mathrm{e}}\right\rangle \otimes|\Phi_{\mathrm{p}}\rangle = \left|\Phi_{\mathrm{e}}\right\rangle \otimes\left(\sum_{n} C_{n} \left|n_{\mathrm{p}}\right\rangle\right).\label{PhotonCI}
\end{equation}
The electronic orbitals and photonic expansion coefficients $C_n$ are to be determined by minimizing the energy
\begin{equation}
	E_{\text{cQED-HF}}=\langle \Psi_0 | \hat{H}_{\text{PF}} | \Psi_0 \rangle.
\end{equation}
However, the expansion with respect to PNs converges extremely slowly in the strong-coupling regime and any truncation would result
in spurious behaviors, e.g., dependence of the energy on the photon
frequency \(\omega\) or violation of translational invariance in charged systems\cite{QEDReview} (see also Sec. \ref{Sec:origin}).
The way out is to go to the coherent-state (CS) representation, which amounts to rotating the reference state $|\Phi_{\mathrm{e}}\rangle\otimes|0_{\mathrm{p}}\rangle$
by the following unitary transformation\cite{QED_HF_CCSD}
\begin{equation}
	\hat{U}_{\mathrm{CS}}=\exp \left(z\left(\hat{b}^{\dagger}-\hat{b}\right)\right), \label{eqn:CS_trans}
\end{equation}
thereby giving rise to the cQED-HF wavefunction
\begin{equation}
	|\Psi_0\rangle_{\text{CS}} = \hat{U}_{\mathrm{CS}} (|\Phi_{\mathrm{e}}\rangle\otimes|0_{\mathrm{p}}\rangle).  \label{eqn:CS_ansatz}
\end{equation}
This is equivalent to applying the unitary transformation \(\hat{U}_{\mathrm{CS}}\) to the PF Hamiltonian \eqref{eqn:generic_elec_boson_h}
\begin{equation}
	\begin{aligned}
		\hat{H}_{\mathrm{CS}} & =\hat{U}_{\mathrm{CS}}^\dag \hat{H}_{\mathrm{PF}} \hat{U}_{\mathrm{CS}} \\
		& =\hat{H}_{\mathrm{PF}}+\omega z^2+\sum_{p q} 2 z g_{p q} E_{pq}+\omega z\left(b+b^{\dagger}\right),
	\end{aligned} \label{eqn:CS_transformed_Hmat}
\end{equation}
followed by the calculation of the energy as
\begin{equation}
	\begin{split}
		E & =   (\langle\Phi_{\mathrm{e}} |\otimes\langle 0_{\mathrm{p}} |)\hat{H}_{\text{CS}}(|\Phi_{\mathrm{e}}\rangle\otimes|0_{\mathrm{p}}\rangle) \\
		  & = \langle \hat{H}_e +  \hat{H}_{\text{DSE}}\rangle + \omega \left(z - \frac{\langle \hat{d} \rangle}{\sqrt{2\omega}}\right)^2 - \frac{\langle\hat{d}\rangle^2}{2},
	\end{split} \label{eqn:E_CS_ansatz}
\end{equation}
where $\langle \hat{O} \rangle$ is a shorthand for $\langle\Phi_{\mathrm{e}} | \hat{O} | \Phi_{\mathrm{e}}\rangle$.
The parameter $z$ in Eq.~\eqref{eqn:CS_trans} can be determined by the stationarity condition $\frac{\partial E}{\partial z}=0$, which can be read directly from Eq.~\eqref{eqn:E_CS_ansatz}:
 \begin{equation}
 	z=\frac{\langle\hat{d}\rangle}{\sqrt{2 \omega}}, \label{eqn:CS_trans_z}
 \end{equation}
which is just a shift of the origin of the harmonic oscillator when $\langle\hat{d}\rangle$ is nonzero.
It can be shown that Eq. \eqref{eqn:CS_ansatz} agrees with Eq. \eqref{PhotonCI} if the expansion goes to infinity in the latter\cite{QED_HF_CCSD}.

Starting from the cQED-HF reference  \eqref{eqn:CS_ansatz} in the CS representation,
various methods can be constructed to account for the electron–electron and electron–photon correlations.
However, it should be noted that, although the cQED-HF total energy is origin-invariant for both neutral and charged systems, the molecular orbitals and the Fock matrix remain origin-dependent for charged species\cite{riso2022molecular,QEDReview}, which in turn makes any subsequent correlation treatment origin-dependent.
One way to circumvent this difficulty is to start from a reference that is fully origin-invariant\cite{riso2022molecular}.
Another route is to search for the exact solution of the PF Hamiltonian for a given (incomplete) one-particle electronic basis,
i.e., cQED full configuration interaction (cQED-FCI). A general cQED-CI ansatz can be written in the PN representation as
\begin{equation}
	|\Psi\rangle = \sum_n \sum_I C_{I\mu, n}\, |I\mu\rangle \otimes |n_{\mathrm{p}}\rangle,
	\label{eqn:cqed_ci_ansatz}
\end{equation}
where the key difference from a conventional CI ansatz for the electronic Hamiltonian is that each term in the QED-CI expansion is a tensor product of a many-electron state \(|I\mu\rangle\) and a photonic state \(|n_{\mathrm{p}}\rangle\).
The number of electrons is conserved, whereas the number of photons can vary.
cQED-FCI includes all possible many-electron states for a given electron number and spin quantum number, and the PN is in principle infinite.
All other wavefunction methods should be regarded as approximations to the cQED-FCI wavefunction. For instance,
the family of cQED-CI methods\cite{QEDReview,QED_CASCI,QED_CIS} approximates the cQED-FCI wavefunction by
truncating both the electronic excitation level and the maximum PN.
In this work, we extend iCIPT2\cite{iCI,iCIPT2,iCIPT2New}
to the regime of strong light–matter coupling, so as to provide a near-cQED-FCI method.

\subsection{Origin Problem}\label{Sec:origin}
Both the coupling and DSE terms of the PF Hamiltonian $\hat{H}_{\mathrm{PF}}$ \eqref{eqn:PF_Hmat} involves the dipole of the molecule.
If the origin $\mathbf{R}$ of the coordinate system is shifted by $\Delta\mathbf{R}$ to $\mathbf{R}^\prime$, the locations of the nuclei and the centers
of the molecular orbitals are shifted by the same amount. Consequently,
\begin{equation}
	\begin{split}
		\bm{\mu}_{p^\prime q^\prime} & = \bigl\langle \phi_p^\prime \big| \bigl( \mathbf{r} - \frac{\sum_I Z_I \mathbf{R}_I^\prime}{N_{\text{e}}} \bigr) \big| \phi_q^\prime \bigr\rangle \\
		& = \bigl\langle \phi_p \big| \bigl( \mathbf{r} + \Delta\mathbf{R} - \frac{\sum_I Z_I (\mathbf{R}_I+\Delta\mathbf{R})}{N_{\text{e}}} \bigr) \big| \phi_q \bigr\rangle \\
		& = \bm{\mu}_{p q} + (1-\frac{ \sum_I Z_I}{N_{\text{e}}}) \Delta\mathbf{R} \delta_{p,q}.
	\end{split}
	\label{eqn:dipole_int2}
\end{equation}
It is clear that the dipole matrix elements and hence $\hat{H}_{\mathrm{PF}}$
 are origin-invariant only when the molecule is neutral, where $N_{\text{e}}=\sum_I Z_I$.
In contrast, for a charged molecule, $\hat{H}_{\mathrm{PF}}$  will depend on the choice of origin. Since the rigid translation of the molecule
is generated by the unitary operator $\hat{U} = e^{i \Delta\mathbf{R} \cdot \hat{\mathbf{p}}}$,
where $\hat{\mathbf{p}}$ is the total momentum operator for the molecular degrees of freedom, $\hat{H}_{\mathrm{PF}}$  for the shifted molecule
can be expressed as $\hat{U}^{\dagger} \hat{H}_{\mathrm{PF}} \hat{U}$, which has the same spectrum as the unshifted $\hat{H}_{\mathrm{PF}}$.
However, this holds only in the limit of a complete photon-number (PN) basis (because $\hat{H}_{\mathrm{PF}}$
does not conserve the PN). A direct deduction is that the energy of a charged molecule will not be
origin-invariant when a finite PN basis is used, which should be avoided in one way or another.
On the other hand, the shift of origin does not modify the purely electronic part of $\hat{H}_{\mathrm{PF}}$,
so that the use of a finite electronic one-particle basis would not be problematic
in the context of origin-invariance.

The situation is different for the CS-transformed Hamiltonian $\hat{H}_{\mathrm{CS}}$ \eqref{eqn:CS_transformed_Hmat}.
When the parameter $z$ in $\hat{U}_{\mathrm{CS}}$ \eqref{eqn:CS_trans} is fixed by the condition \eqref{eqn:CS_trans_z},
$\hat{H}_{\mathrm{CS}}$ can be written as
\begin{equation}
	\begin{aligned}
		\hat{H}_{\mathrm{CS}} = &\ \hat{H}_{\mathrm{e}} + \omega \hat{b}^{\dagger} \hat{b}
		- \sqrt{\frac{\omega}{2}} \bigl[ \hat{d}_{\mathrm{e}} - \langle \hat{d}_{\mathrm{e}} \rangle \bigr] \bigl( \hat{b}^{\dagger} + \hat{b} \bigr) +  \frac{1}{2} \bigl[ \hat{d}_{\mathrm{e}} - \langle \hat{d}_{\mathrm{e}} \rangle \bigr]^{2},
	\end{aligned}
	\label{eqn:CS_trans_Hmat2}
\end{equation}
where $\langle \hat{d}_{\mathrm{e}} \rangle = \bm{\lambda} \cdot \langle \hat{\boldsymbol{\mu}}_{\mathrm{e}} \rangle$.
Although the parameter $z$ depends on the displacement $\Delta\mathbf{R}$, $\hat{H}_{\mathrm{CS}}$ \eqref{eqn:CS_trans_Hmat2} does not: $\Delta\mathbf{R}$ contributes equally to $\langle \hat{d}_{\mathrm{e}} \rangle$ and $\hat{d}_{\mathrm{e}}$ and thus cancels each other.
Consequently, $\hat{H}_{\mathrm{CS}}$ is origin‑invariant. 

\section{cQED-iCIPT2}\label{Sec-cQED-iCIPT2}
\subsection{iCIPT2}
Like other variants of sCIPT2, iCIPT2 involves three key steps:
iterative update of the variational space $P$ from a guess space $P_0$  (selection step), diagonalization and pruning of
the variational space (diagonalization step),
and a second-order perturbation correction to the energy (perturbation step).
Starting from the guess wave function $|\Psi^{(0)}\rangle=\sum_P C_{J\nu} | J\nu\rangle$ and energy $E_{\text{var}}^{(0)}$,
the following ``iCI criterion'' characterized by the boolean function $f(|I\mu\rangle, |J\rangle, C_{\mathrm{min}})$ is adopted
in iCIPT2 to select important CSFs:
\begin{itemize}
	\item [(A)] If $|I\rangle$ is identical with or singly excited from $|J\rangle$, then
	\begin{equation}
		\begin{aligned}
			& f\left(|I \mu\rangle,|J\rangle, C_{\min }\right)=\left(\max _\nu\left(\left|H_{\mu \nu}^{I J} C_{J\nu}\right|\right) \geq C_{\min }\right) \quad \text { and } \\
			& \left(\max _\nu\left(\left|\frac{H_{\mu \nu}^{I J} C_{J\nu}}{E_{\text{var}}^{(0)}-H_{\mu \mu}^{I I}}\right|\right) \geq C_{\min }\right);
		\end{aligned} \label{eqn:selection1}
	\end{equation}
	\item [(B)] If $|I\rangle$ is doubly excited from $|J\rangle$, then
	\begin{equation}
		\begin{aligned}
			& \left.f(| I \mu\rangle,|J\rangle, C_{\min }\right)=\left(\max _\nu\left(\left|\tilde{H}^{I J} C_{J\nu}\right|\right) \geq C_{\min }\right) \quad \text { and } \\
			& \quad\left(\max _\nu\left(\left|H_{\mu \nu}^{I J} C_{J\nu}\right|\right) \geq C_{\min }\right) \quad \text { and } \\
			& \quad\left(\max _\nu\left(\left|\frac{H_{\mu \nu}^{I J} C_{J\nu}}{E_{\text{var}}^{(0)}-H_{\mu \mu}^{I I}}\right|\right) \geq C_{\min }\right),
		\end{aligned}\label{eqn:selection2}
	\end{equation}
\end{itemize}
where $|J\rangle \in P_0$ and $|I\mu\rangle\in Q$, with $Q$ being the first-order interacting space (FOIS) of $P_0$.
The newly selected CSFs are added to $P_0$ to form an expanded space $P^\prime$ (cf. Algorithm \ref{alg:selection}), on which the Hamiltonian is constructed and then diagonalized.
Those CSFs with coefficients smaller than $C_{\text{min}}$ in absolute value are discarded, resulting in an updated prime space $P$.
The procedure is repeated until the similarity
$\frac{\left|P_0 \cap P\right|}{\left|P_0\cup P\right|}$
between $P_0$ and $P$ of two adjacent iterations has reached a threshold (typically 95\%).

The above procedure generates a very compact wavefunction within a few cycles (see Ref. \citenum{iCIPT2New} for comparison with other criteria).
The residual dynamic correlation is then estimated with ENPT2:
\begin{equation}
	\begin{aligned}
		E_{c}^{(2)}=&\sum_{|I \mu\rangle \in Q} \frac{\left|\left\langle I \mu|H| \Psi^{(0)}\right\rangle\right|^2}{E^{(0)}_{\text{var}}-H_{\mu \mu}^{I I}} \\
		=&\sum_{|I \mu\rangle \in Q} \frac{\left|\sum_{|J \nu\rangle \in P} H_{\mu \nu}^{I J} C_{J\nu}\right|^2}{E^{(0)}_{\text{var}}-H_{\mu \mu}^{I I}}. \label{pt2_formula}
	\end{aligned}
\end{equation}
Here, $H^{IJ}_{\mu\nu} = \langle I\mu | H | J\nu\rangle$ is the Hamiltonian matrix elements between CSFs $|I\mu\rangle$ and $|J\nu\rangle$.
A pipelined, residue- and constrained-based\cite{ASCI2018PT2} algorithm has been designed\cite{MetaWave} to evaluate Eq. \eqref{pt2_formula} efficiently.
It involves systematic construction of the FOIS $Q$ and identification of the interacting orbital configuration (oCFG) pairs $(I,J)$ with $I\in P$ and $J\in Q$, as well as the tabulated unitary group approach (TUGA) \cite{iCIPT2,iCIPT2New} for evaluating the numerator and denominator in Eq.~\eqref{pt2_formula}.

iCIPT2 can be readily combined with the PF Hamiltonian \eqref{eqn:PF_Hmat}/\eqref{eqn:generic_elec_boson_h} or its CS-transformed
form \eqref{eqn:CS_transformed_Hmat}, leading to cQED-PN-iCIPT2 and cQED-CS-iCIPT2, respectively  (see Sec. \ref{sec:impl}).

\begin{algorithm}[H]
	\caption{Selection Algorithm in iCIPT2 }\label{alg:selection}
	\begin{algorithmic}[1]
		\Require Variational space $P_0$, wavefunction $|\Psi^{(0)}\rangle = \sum C_{J\nu}|J\nu\rangle$, threshold $C_{\text{min}}$
		\Ensure  New variational space $P^\prime$
		\State $P^\prime\gets \{\}$
		\For {$|J\nu\rangle$ in $P_0$}
		\State $S_{J\nu}\gets$ candidate CSFs in $Q$ via Eqs. \eqref{eqn:selection1} and \eqref{eqn:selection2}
		\State $P^\prime\gets P^\prime\bigcup S_{J\nu}$
		\EndFor
		\State \Return $P^\prime$
	\end{algorithmic}
\end{algorithm}
\subsection{Extension of iCIPT2 to cQED}\label{sec:impl}
To adapt iCIPT2 to the tensor-product structure of the cQED-CI ansatz \eqref{eqn:cqed_ci_ansatz}, we first introduce a
graded configuration space in Sec.~\ref{sec:graded_space}.
After presenting the necessary formulas for the matrix elements in Sec.~\ref{sec:matrix_elmts}, we detail
the adaptation of the key steps of iCIPT2 to the graded configuration space in Sec.~\ref{sec:algo}. We then discuss the use of symmetry
in Sec. \ref{sec:symm} and the energy extrapolation in \ref{sec:extra}.
The origin dependence of cQED-iCIPT2 is discussed in Sec.~\ref{sec:origin_dependence},
where we demonstrate that, while the (near-exact) total energy remains origin-invariant, the compactness of the wavefunction is affected by the choice of origin in cQED-PN-iCIPT2 but not in cQED-CS-iCIPT2.

\subsubsection{Graded Configuration Space}\label{sec:graded_space}
In mathematics, a graded algebraic structure (e.g., ring, module, or vector space) is one that decomposes into a direct sum of components indexed by integers.
In the context of cQED, the Fock space for a single cavity mode is naturally graded by the PN:
\begin{equation}
	F = \bigoplus_{n \geq 0} F_n, \label{eqn:graded_space}
\end{equation}
where $F_n$ is the subspace spanned by basis states of
the form $|I\mu\rangle \otimes |n_{\textrm{P}}\rangle$ with a fixed PN $n$ (and a fixed electron number $N_{\mathrm{e}}$ as in standard CI).
For a multimode cavity, the Fock space is graded by the vector of occupation numbers of each mode.
Owing to the form of the PF Hamiltonian and its CS-transformed form,
subspaces whose PNs differ by two or more are uncoupled, so that matrix elements in between vanish identically.
We note that in mathematics the term ``graded'' also carries a multiplicative aspect: the product of an element of degree $m$ with an element of degree $n$ has degree $m+n$.
In the present work, however, we use the term only in the additive sense, i.e., to organize the Fock space into a direct sum of fixed-PN subspaces.

\subsubsection{cQED Hamiltonian Matrix Elements} \label{sec:matrix_elmts}
The cQED Hamiltonian matrix elements are denoted as $\langle I^n_\mu | \hat{H} | J^m_\nu \rangle$, where
 $|I^n_\mu\rangle = |I\mu\rangle \otimes |n_{\mathrm{P}}\rangle$.
According to the graded Fock-space decomposition \eqref{eqn:graded_space},
the matrix elements fall into two categories:
intra-subspace ($n=m$) and inter-subspace  ($n\neq m$). For the former,
the non-vanishing matrix elements of the generic Hamiltonian \eqref{eqn:generic_elec_boson_h}
 arise only between oCFG pairs $(I,J)$ that differ by at most two electrons, and
only the first three terms of Eq. \eqref{eqn:generic_elec_boson_h} have contributions.
The expressions are hence identical to those of the Born–Oppenheimer electronic Hamiltonian\cite{iCIPT2,iCIPT2New},
except that the diagonal elements should be shifted by $n\omega$.
Therefore, the existing infrastructure of MetaWave\cite{MetaWave} can directly be reused here.
As for the $\hat{H}_{\mathrm{CS}}$ Hamiltonian \eqref{eqn:CS_transformed_Hmat},
the term $\omega z^2$ provides a common energy shift for all states,
while the third term therein acts as a one-body operator and
can be absorbed by modifying the one-electron integrals as $h_{pq}' = h_{pq} + 2z g_{pq}$.
As such, the matrix elements reduce again to those of the
Born–Oppenheimer electronic Hamiltonian\cite{iCIPT2,iCIPT2New}, with the diagonal elements shifted by $n\omega$.

As for the inter-subspace case, only the $\hat{h}_{\text{e-p}}$ term of the generic Hamiltonian \eqref{eqn:generic_elec_boson_h}
has non-vanishing matrix elements
\begin{equation}
	\begin{split}
		\langle I_\mu^n|\hat{h}_{\text{e-p}}|J_\nu^m \rangle & = \langle I_\mu^n| \sum_{pq} g_{pq} E_{pq} \left(b+b^\dagger\right)  |J_\nu^m \rangle \\
		& = \sum_{pq} g_{pq}\langle I\mu |E_{pq}|J\nu\rangle \left(\langle n_{\textrm{P}}|b|m_{\textrm{P}}\rangle+\langle n_{\textrm{P}}|b^\dagger|m_{\textrm{P}}\rangle\right) \\
		& = \sum_{pq} g_{pq}\langle I\mu |E_{pq}|J\nu\rangle \left(\sqrt{m}\delta_{n,m-1}+\sqrt{m+1}\delta_{n,m+1}\right).
	\end{split} \label{eqn:inter_hmat_1}
\end{equation}
Yet, the $\hat{H}_{\mathrm{CS}}$ Hamiltonian \eqref{eqn:CS_transformed_Hmat} has an additional contribution from the last term therein
\begin{equation}
	\begin{split}
		\langle I_\mu^n|\omega z(b+b^\dagger)|J_\nu^m \rangle & = \omega z \delta_{I,J}\delta_{\mu,\nu}\left(\langle n_{\textrm{P}}|b|m_{\textrm{P}}\rangle+\langle n_{\textrm{P}}|b^\dagger|m_{\textrm{P}}\rangle\right) \\
		& = \omega z \delta_{I,J}\delta_{\mu,\nu} \left(\sqrt{m}\delta_{n,m-1}+\sqrt{m+1}\delta_{n,m+1}\right).
	\end{split} \label{eqn:inter_hmat_2}
\end{equation}

\subsubsection{Implementation}\label{sec:algo}

Having established the background, it remains to see how to adapt the key steps of iCIPT2 to cQED-iCIPT2 calculations.
The key observation is that, in the context of graded configuration space,
all steps of iCIPT2 can be organized as a loop over pairs of connected subspaces $(P_i, W_j)$, where $P_i$ and $W_j$
belong to spaces $P$ and $W$, respectively.
Here, $P$ is the primary space and $W$ is either $P$ itself (for Hamiltonian construction and matrix-vector product (MVP)) or
the FOIS $Q$ of $P$ (for selection and ENPT2).
As discussed previously, only pairs of subspaces with the same PN or with PNs differing by $\pm 1$ have non-vanishing matrix elements.

Consider first the selection step, where $W=Q$. The algorithm for this step  (see Algorithm \ref{alg:cqed_selection}) is simply a nested Algorithm \ref{alg:selection}.
That is, Algorithm \ref{alg:selection} can directly be used for each pair of connected subspaces $(P_i, Q_j)$.
It is just that the matrix elements required by Eq. \eqref{eqn:selection1} should be calculated
as described in Sec. \ref{sec:matrix_elmts}.

The ENPT2 step can be adapted in a similar manner and is hence not repeated here.

\begin{algorithm}[H]
	\caption{Selection Algorithm in cQED-iCIPT2}\label{alg:cqed_selection}
	\begin{algorithmic}[1]
		\Require Variational space $P_0 = \bigoplus_{i\geq 0} P_{0,i}$, wavefunction $|\Psi^{(0)}\rangle = \sum_{i\geq 0}\sum_{J\nu} C_{J\nu,i} |J\nu\rangle \otimes |i_{\mathrm{P}}\rangle$, threshold $C_{\text{min}}$
		\Ensure  New variational space $P^\prime=\bigoplus_{i\geq 0} P_i^\prime$
		\State Let $\mathcal{I} = \{ i \mid P_{0,i} \text{ is non-empty} \}$
		\State Construct list $A \gets \{ (i,j) \mid i \in \mathcal{I},\; j \in \{i,\, i+1,\, i-1\} \}$
		\For {$(i,j)$ in $A$}
		\State $P_j^\prime\gets \{\}$ \Comment{Initialize all required subspaces}
		\EndFor
		\For {$(i,j)$ in A}
		\For {$|J\nu\rangle$ in $P_{0.i}$}
		\State $S_{J\nu}\gets$ candidate basis in $Q_j$ via Eqs. \eqref{eqn:selection1} and \eqref{eqn:selection2}
		\State $P^\prime_j\gets P^\prime_j\bigcup S_{J\nu}$
		\EndFor
		\EndFor
		\State $P^\prime\gets\bigoplus_{i\geq 0} P_i^\prime$
		\State \Return $P^\prime$
	\end{algorithmic}
\end{algorithm}

The diagonalization step amounts to finding the ground state iteratively (via the iterative vector interaction
(iVI) approach\cite{iVI}).
In MetaWave, iVI is implemented as a C++ class
that accepts the MVP as another class.
Thus, for different Hamiltonians, only the MVP implementation needs to be supplied.
The Hamiltonian matrix elements are first organized into intra- or inter-subspace blocks according to the PNs in the bra and ket.
For each block, the corresponding sigma vector is updated, accounting for the Hermiticity (see Algorithm~\ref{alg:mvp}).

\begin{algorithm}[H]
	\caption{Matrix‑Vector Product Algorithm in cQED-iCIPT2}\label{alg:mvp}
	\begin{algorithmic}[1]
		\Require Variational space $P = \bigoplus_{i\geq 0} P_i$, wavefunction $|\Psi^{(0)}\rangle = \sum_{i\geq 0}\sum_{J\nu} C_{J\nu,i} |J\nu\rangle \otimes |i_{\mathrm{P}}\rangle$
		\Ensure Sigma vector $\sigma = \hat{P} H \hat{P} |\Psi^{(0)}\rangle$
		\State Let $\mathcal{I} = \{ i \mid P_i \text{ is non-empty} \}$
		\State Construct list $A \gets \{ (i,j) \mid i,j \in \mathcal{I},\; j \in \{i,\, i+1\} \}$ \Comment{Hermiticity is taken into consideration}
		\State Initialize subspace sigma vectors $\sigma_i \gets \mathbf{0}$ for all $i\geq 0$.
		\For {$(i,j)$ in $A$}
		\If{$i=j$} \Comment{Intra-subspace case}
		\For{each matrix element in sub-block $\mathbf{H}^{i,i}$}
		\State $|I\mu\rangle\gets$ bra index
		\State $|J\nu\rangle\gets$ ket index
		\If{$|I\mu\rangle=|J\nu\rangle$}
		\State $\sigma_{I\mu,i} \, +\!\!= (H^{i,i})^{II}_{\mu\mu} C_{I\mu,i}$
		\Else \Comment{Hermiticity is taken into consideration}
		\State $\sigma_{I\mu,i} \, +\!\!= (H^{i,i})^{IJ}_{\mu\nu} C_{J\nu,i}$
		\State $\sigma_{J\nu,i} \, +\!\!= \bigl[(H^{i,i})^{IJ}_{\mu\nu}\bigr]^* C_{I\mu,i}$
		\EndIf
		\EndFor
		\Else \Comment{Inter-subspace case}
		\For{each matrix element in sub-block $\mathbf{H}^{i,j}$}
		\State $|I\mu\rangle\gets$ bra index
		\State $|J\nu\rangle\gets$ ket index
		\State $\sigma_{I\mu,i} \, +\!\!= (H^{i,j})^{IJ}_{\mu\nu} C_{J\nu,j}$
		\State $\sigma_{J\nu,j} \, +\!\!= \bigl[(H^{i,j})^{IJ}_{\mu\nu}\bigr]^* C_{I\mu,i}$ \Comment{Hermiticity is taken into consideration}
		\EndFor
		\EndIf
		\EndFor
		\State $\sigma \gets \bigoplus_{i\geq 0} \sigma_i$
		\State \Return $\sigma$
	\end{algorithmic}
\end{algorithm}

\subsubsection{$\mathbb{Z}_2$ Symmetry} \label{sec:symm}
The proper use of symmetries—whether spatial or spin—can significantly reduce computational cost.
Electron spin remains a good quantum number for the (nonrelativistic) cQED Hamiltonian and is explicitly utilized in iCIPT2 by adopting CSFs as the many-electron basis.
The spatial symmetry of the system, however, is lowered by the bilinear coupling $\hat{H}_{\text{int}}$ and the DSE $\hat{H}_{\text{DSE}}$.
This can be attributed to the external electric field, which breaks the spatial symmetry of the isolated molecule.
Although spatial symmetry may be retained when the cavity photon is polarized along certain directions, we do not exploit this feature in the present work.
Nevertheless, for molecules that are invariant under spatial inversion, a hidden $\mathbb{Z}_2$ symmetry persists.
This follows from the observation that the dipole operator $\hat{\mathbf{d}} = \bm{\lambda}\cdot\hat{\boldsymbol{\mu}}$ transforms as the $A_u$ irreducible representation (irrep) of the $C_i$ point group.
Consequently, the DSE term $\hat{H}_{\text{DSE}} \propto (\bm{\lambda}\cdot\hat{\boldsymbol{\mu}})^2$ belongs to the totally symmetric $A_g$ irrep.
The coupling term $\hat{H}_{\text{int}}$ transforms as $A_u$ but, in addition, changes the PN by $\pm 1$.
All remaining terms ($\hat{H}_{\text{e}}$ and $\hat{H}_{\text{cav}}$) are fully symmetric and conserve the PN.
We may therefore assign the label $0$ to $A_g$ and $1$ to $A_u$, and define
\begin{equation}
	f(|I_\mu^n\rangle) = \bigl( \text{irrep label of } |I\mu\rangle + n \bigr) \bmod 2
\end{equation}
for a basis state $|I_\mu^n\rangle = |I\mu\rangle \otimes |n_{\mathrm{P}}\rangle$.
The cQED Hamiltonian does not change the value of $f$, so that every eigenstate can be classified by $f \in \{0,1\}$.
For example, the two states $|I\mu\rangle \otimes |0_{\mathrm{P}}\rangle$ (with $|I\mu\rangle \in A_g$) and $|J\nu\rangle \otimes |0_{\mathrm{P}}\rangle$ (with $|J\nu\rangle \in A_u$) carry $f = 0$ and $f = 1$, respectively, and cannot mix.
We refer to this as $\mathbb{Z}_2$ symmetry for inversion-symmetric molecules in an optical cavity.
Exploiting this symmetry reduces the computational cost by a factor of two.
To the best of our knowledge, this symmetry is identified here for the first time in the context of cQED.

\subsubsection{Extrapolation} \label{sec:extra}
\begin{figure}[!ht]
	\includegraphics[width=1.0\textwidth]{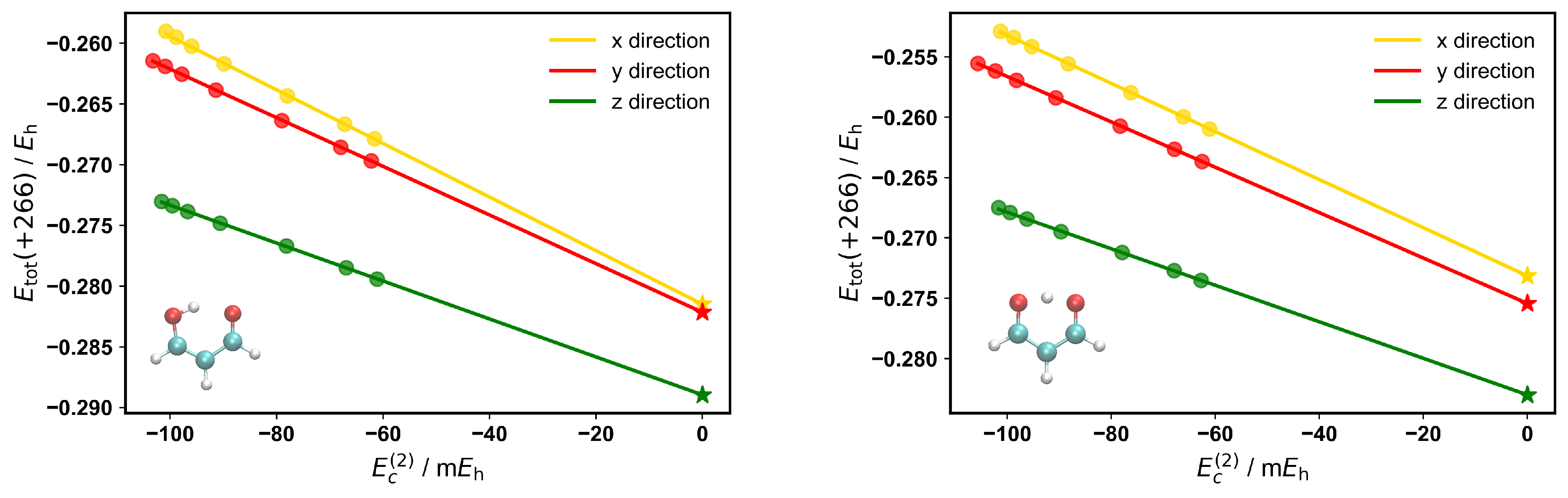}
	\caption{
		cQED-CS-iCIPT2/cc-pVDZ calculations of the proton transfer reaction in malonaldehyde at the reactant (left panel) and the transition state geometry (right panel) inside an optical cavity with cavity parameters $\omega = 3$ eV and $|\lambda|=0.1$ au.
		Three possible orientations of the coupling vectors are tested.
		}
	\label{fig:extra}
\end{figure}

A single parameter, $C_{\text{min}}$, is invoked in iCIPT2 to control the size of
the variational space $P$ and the corresponding variational energy $E_{\text{var}}$.
The ENPT2 correction $E_{\text{c}}^{(2)}$ is then computed for this space, yielding the total energy $E_{\text{tot}} = E_{\text{var}} + E_{\text{c}}^{(2)}$.
As is common in sCIPT2, for a sufficiently large variational space, $E_{\text{tot}}$ scales linearly with $E_{\text{c}}^{(2)}$,
so that a linear extrapolation to the limit $E_{\text{c}}^{(2)} \rightarrow 0$ recovers the FCI limit, making iCIPT2 a near-exact method.
This behavior is also observed in the context of cQED-CS-iCIPT2 (in conjunction with the DSE \eqref{eqn:dse_sp_form}),
as shown in Fig.~\ref{fig:extra} for the proton transfer reaction of
malonaldehyde at the reactant and transition state geometries along with three orientations for the coupling vector.
In the remainder of this paper, we extract the FCI energy by performing a weighted linear extrapolation with weight $1/|E_{\text{c}}^{(2)}|$.

\subsubsection{Origin Dependence of cQED-iCIPT2} \label{sec:origin_dependence}


Since the $\hat{H}_{\mathrm{CS}}$ Hamiltonian is origin-invariant,
cQED-CS-iCIPT2 is naturally origin-invariant even when an incomplete PN basis is used. In contrast,
cQED-PN-iCIPT2 based on the PF Hamiltonian $\hat{H}_{\mathrm{PF}}$ becomes origin-invariant for charged systems
only when a nearly complete PN basis is adopted. Although this is guaranteed by the selection,
an improper choice of origin can lead to a very delocalized wavefunction, so as to slow down the selection.
To see this, we first introduce the following relation
\begin{equation}
	|\Psi_0\rangle_{\text{PN}} = \exp\!\left[-\frac{\langle \hat{d}\rangle}{\sqrt{2\omega}} (\hat{b}^\dagger - \hat{b})\right] |\Psi_0\rangle_{\text{CS}}
	\approx \exp\!\left[-\frac{\langle \hat{d}\rangle}{\sqrt{2\omega}} (\hat{b}^\dagger - \hat{b})\right] \bigl(|\Psi_0^{\mathrm{e}}\rangle \otimes |0_{\mathrm{P}}\rangle \bigr), \label{eqn:origin_dependence_ansatz}
\end{equation}
where $|\Psi_0^{\mathrm{e}}\rangle$ is the electronic part of $|\Psi_0\rangle_{\text{CS}}$
 obtained by projecting onto the photon vacuum.
Eq.~\eqref{eqn:origin_dependence_ansatz} assumes that the ground state $|\Psi_0\rangle_{\text{CS}}$ of
$\hat{H}_{\mathrm{CS}}$ is dominated by the zero‑photon component. Tracing out the electronic degrees of freedom
from both sides of Eq. \eqref{eqn:origin_dependence_ansatz} then gives rise to the following relation for the photonic reduced density matrix
	\begin{equation}
		\operatorname{Tr}_e(_{\mathrm{PN}}|\Psi_0\rangle\langle\Psi_0|_{\mathrm{PN}}) = \sum_{n\geq 0} P(n) |n_P\rangle\langle n_P|, \quad
		P(n) = \exp\!\left(-\frac{\langle \hat{d} \rangle^2}{2\omega}\right) \frac{\langle \hat{d} \rangle ^{2n}}{(2\omega)^n}\frac{1}{n!},
	\end{equation}
	which is a Poisson distribution $P(n)=e^{-\lambda}\frac{\lambda^n}{n!}$ with $\lambda = \langle \hat{d} \rangle^2 / 2\omega$.
It follows that $P(0)=\exp(-\lambda)$ can be made arbitrarily small and both the mean and variance of the PN (equal to $\lambda$) can
be made arbitrarily large via the shift of origin. Consequently,
the weights $|c_{I\mu,i}|^2$ can spread broadly over many PN states rather than centering around the zero-photon component.
This will render the convergence of the selection step very slow, even though cQED‑PN-iCIPT2 can determine the proper PN automatically.

\begin{figure}[!ht]
	\includegraphics[width=0.9\textwidth]{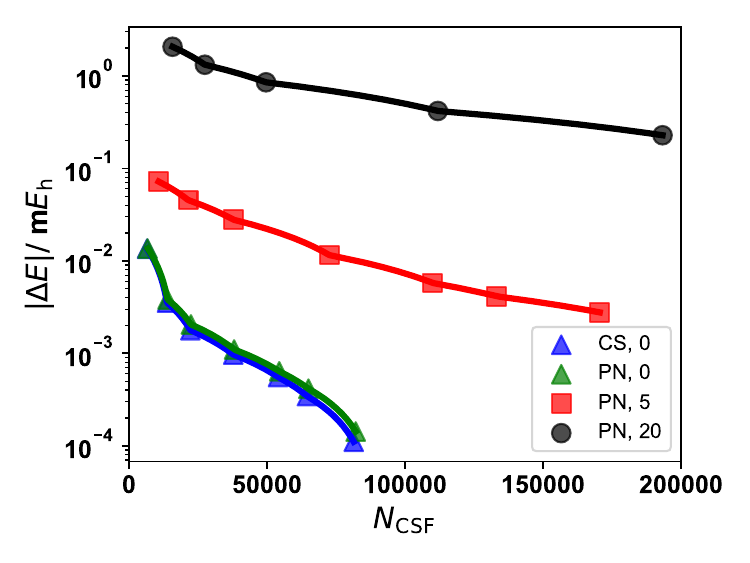}
	\caption{Absolute energy errors of cQED-CS/PN-iCIPT2 for \ce{OH-}/cc-pVDZ
		with respect to the number of polaritonic configuration state functions ($N_{\text{CSF}}$).
		\ce{OH-} is placed  along the $x$ axis with \ce{O} and \ce{H} at $x_{\text{O}}$ and $x_{\text{O}}+0.9$~\AA~respectively.
		The calculations employ cavity parameters $\omega=5.96$ $\mathrm{eV}$ and $|\lambda|=0.05$ $\mathrm{au}$.
		The proton mode is polarized along the molecule.
		cQED-CS-iCIPT2 is tested only for $x_{\text{O}}=0$, whereas cQED-PN-iCIPT2 is tested for $x_{\text{O}}=0,5$ and 20 (numerical labels in the legend).
	}
	\label{fig:origin_dependence_selection}
\end{figure}

\begin{figure}[!ht]
	\includegraphics[width=0.9\textwidth]{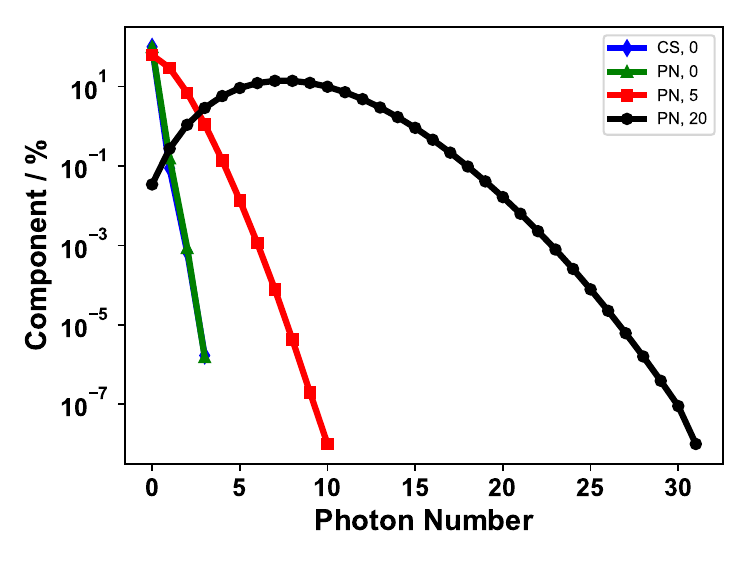}
	\caption{
		Distribution (in \%) of $\sum\left|c_{I\mu,i}\right|^2$ over photon numbers for the cQED-CS/PN-iCIPT2 wavefunctions
of \ce{OH-}/cc-pVDZ with $C_{\text{min}}=5\times 10^{-6}$.
		\ce{OH-} is placed along the $x$ axis with \ce{O} and \ce{H} at $x_{\text{O}}$ and $x_{\text{O}}+0.9$~\AA, respectively.
		The calculations employ cavity parameters $\omega=5.96$ $\mathrm{eV}$ and $|\lambda|=0.05$ $\mathrm{au}$.
		The proton mode is polarized along the molecule.
		cQED-CS-iCIPT2 is tested only for $x_{\text{O}}=0$, whereas cQED-PN-iCIPT2 is tested for $x_{\text{O}}=0,5$ and 20 (numerical labels in the legend).
	}
	\label{fig:origin_dependence_comp}
\end{figure}

To verify the above analysis, we take the hydroxide anion \ce{OH-} with the cc-pVDZ basis set at a bond length of 0.9~\AA{} as an example.
The molecule is placed along the $x$ axis, with \ce{O} and \ce{H} located at $x_{\text{O}}$ and $x_{\text{O}} + 0.9$~\AA, respectively.
The cavity parameters are $\omega = 5.96\ \mathrm{eV}$ and $|\lambda| = 0.05\ \mathrm{au}$, and the photon polarization is aligned with the molecular axis.
For cQED-CS-iCIPT2 we only examine $x_{\text{O}} = 0$~\AA, whereas for cQED-PN-iCIPT2 we test $x_{\text{O}} = 0$, $5$, and $20$~\AA.
To illustrate the influence of different origins on the compactness of the wavefunction,
we plot the energy error (the difference between the total energy and the converged energy) versus the number of polaritonic CSFs ($N_{\text{CSF}}$) in Fig.~\ref{fig:origin_dependence_selection}.
It is clear that as $x_{\text{O}}$ increases, the energy error with a fixed $N_{\text{CSF}}$ also increases.
The distribution (in \%) of $\sum |c_{I\mu,i}|^2$  over the PN index is further
shown in Fig.~\ref{fig:origin_dependence_comp}.
For $x_{\text{O}} = 0$~\AA, the distributions for the cQED-CS-iCIPT2 and cQED-PN-iCIPT2 wavefunctions are nearly identical.
However, for large $x_{\text{O}}$, the cQED-PN-iCIPT2 distribution becomes significantly broader,
to the extent that photon states of PNs larger than 30 still have discernible contributions at $x_{\text{O}} = 20$~\AA.

\clearpage
\newpage

\section{Results and Discussion}\label{sec:pilot_app}

All cQED-HF calculations are performed with the POLAR program\cite{POLAR}.


\subsection{Potential Energy Curves of \ce{N2} in Cavity} \label{sec:n2_curve}

\begin{figure}[!ht]
	\includegraphics[width=0.9\textwidth]{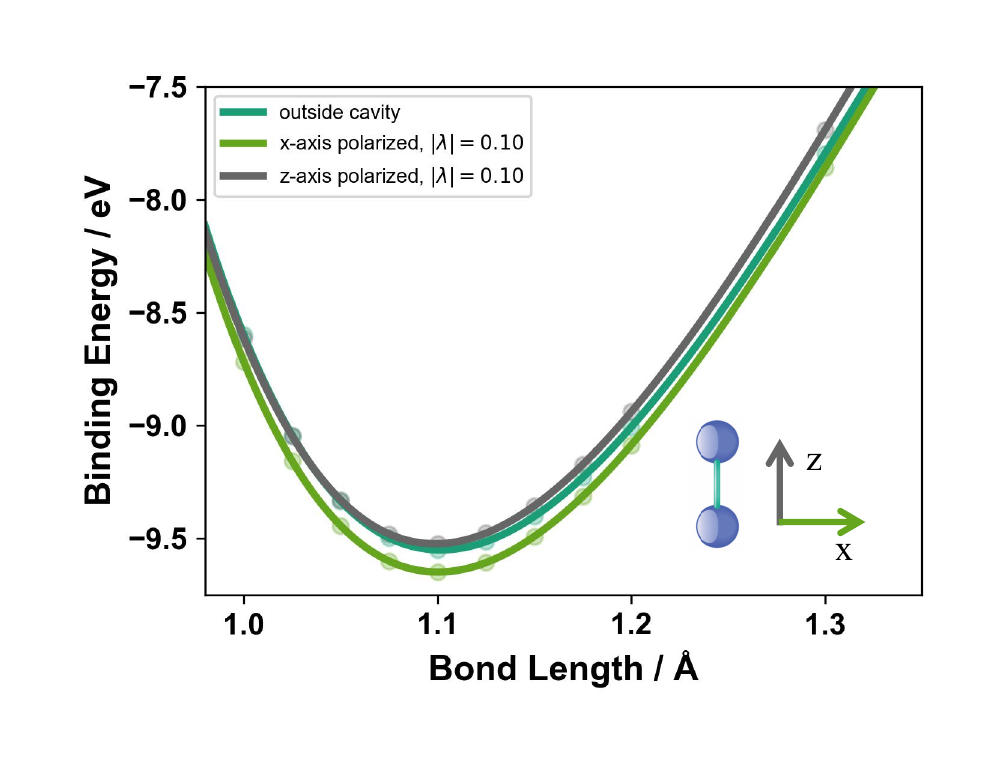}
	\caption{Potential energy curves (in eV) for the
		ground states of \ce{N2} outsize and inside an optical cavity.
		 The cQED-iCIPT2/cc-pVTZ calculations employ cavity parameters $\omega=0.5$ eV and a fixed coupling strength $|\lambda|=0.10$.
		 The coupling vectors are oriented along ($z$-direction) and perpendicular ($x$-direction) to the molecule.
		}
	\label{fig:cqed_n2}
\end{figure}

\begin{figure}[!ht]
	\includegraphics[width=0.9\textwidth]{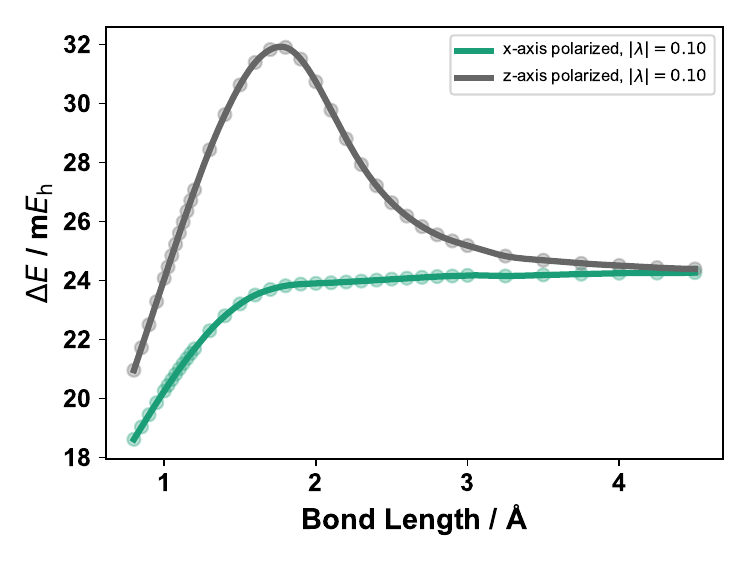}
	\caption{
		Energy difference $\Delta E$ between the ground states of
 \ce{N2} inside and outsize an optical cavity.
		The cQED-iCIPT2/cc-pVTZ  calculations employ cavity parameters $\omega=0.5$ eV and a fixed coupling strength $|\lambda|=0.10$.
		The coupling vectors are oriented along ($z$-direction) and perpendicular ($x$-direction) to the molecule.
	}
	\label{fig:cqed_n2_diff}
\end{figure}

Bond breaking is a typical case that requires a multi-configurational description.
The situation may become more complicated when the molecule is coupled with an optical cavity. To see this,
we perform cQED-iCIPT2/CAS(14e,60o)/cc-pVTZ calculations (with cavity-free CASSCF(6e,6o) orbitals)
on the potential energy curves of \ce{N2} inside a cavity
characterized by $\omega = 0.5$~eV (NB: cQED-CS-iCIPT2 and cQED-PN-iCIPT2 are identical in this case).
Two orientations of the coupling vector are considered: along or perpendicular to the molecular axis.
As shown in Table~\ref{table:n2_curve} and Fig.~\ref{fig:cqed_n2}, the interaction of \ce{N2} with the cavity photon
tends to increase the energy but not uniformly along the distance. As a result,
the equilibrium bond length of \ce{N2} is shortened, which is companied by an enlarged harmonic vibrational frequency
for both polarization directions.
When the cavity mode is polarized along the molecular axis, the energy in the bonding region is raised more strongly (see Fig.~\ref{fig:cqed_n2_diff}), 
resulting in a smaller dissociation energy $D_e$. The larger the coupling strength, the smaller the $D_e$. 
The opposite is true when the polarization is perpendicular to the molecular axis.

\begin{table}[!ht]
	\small
	\centering
	\caption{Equilibrium bond length $r_e$ (in~\AA), harmonic vibrational frequency $\omega_e$ (in cm$^{-1}$) and dissociation energy $D_e$ (in eV) for the ground states of \ce{N2} outside and inside an optical cavity.
	The cQED-iCIPT2/cc-pVTZ calculations employ cavity parameters $\omega=0.5$ eV and different coupling strengths $|\lambda|$.
	The coupling vectors are oriented along ($z$-direction) and perpendicular ($x$-direction) to the molecule.
	}
	\begin{threeparttable}
		\begin{tabular}{c|cccc} \toprule
			& \multicolumn{1}{c}{$|\lambda|$}
			& \multicolumn{1}{c}{$r_e$}
			& \multicolumn{1}{c}{$\omega_e$}
			& \multicolumn{1}{c}{$D_e$} \\ \toprule
			outside cavity
			& \multicolumn{1}{c}{-} & 1.1016 & 2350 & 9.551 \\\midrule
			\multirow{4}{*}{x direction}
			&0.01&1.1016&2349&9.552 \\
			&0.03&1.1015&2350&9.560 \\
			&0.05&1.1013&2353&9.575 \\
			&0.10&1.1002&2359&9.649 \\ \midrule
			\multirow{4}{*}{z direction}
			&0.01&1.1016&2349&9.551 \\
			&0.03&1.1013&2352&9.548 \\
			&0.05&1.1008&2357&9.542 \\
			&0.10&1.0988&2371&9.524 \\ \midrule
		\end{tabular}
	\end{threeparttable} \label{table:n2_curve}
\end{table}

\clearpage
\newpage

\subsection{Potential Energy Curves of \ce{C2H4}} \label{sec:c2h4_curve}

The double-bond torsion of \ce{C2H4} has been the subject of numerous theoretical and experimental studies over the past 50 years\cite{C2H4_Review}.
Because the ground and excited states possess open-shell character, especially at the dihedral angle $\phi_{\text{HCCH}} = 90^\circ$, single-reference methods 
often suffer from severe spin contamination. For instance,  
spin-flip time-dependent density functional theory yields $\text{S}_1$ and $\text{T}_1$ states
that are heavily spin-contaminated and become erroneously degenerate\cite{c2h4_tddft}.
These deficiencies create additional complications when an optical cavity is introduced: the heavily spin-contaminated states exhibit non-vanishing interactions with the $|\text{S}_0\rangle \otimes |1_{\mathrm{P}}\rangle$ state, leading to a qualitatively wrong description of the polaritonic states.
Therefore, an accurate treatment of the low-lying states of \ce{C2H4} both outside and inside an optical cavity requires spin-adapted multi-reference methods.

\begin{figure}[!ht]
	\includegraphics[width=0.9\textwidth]{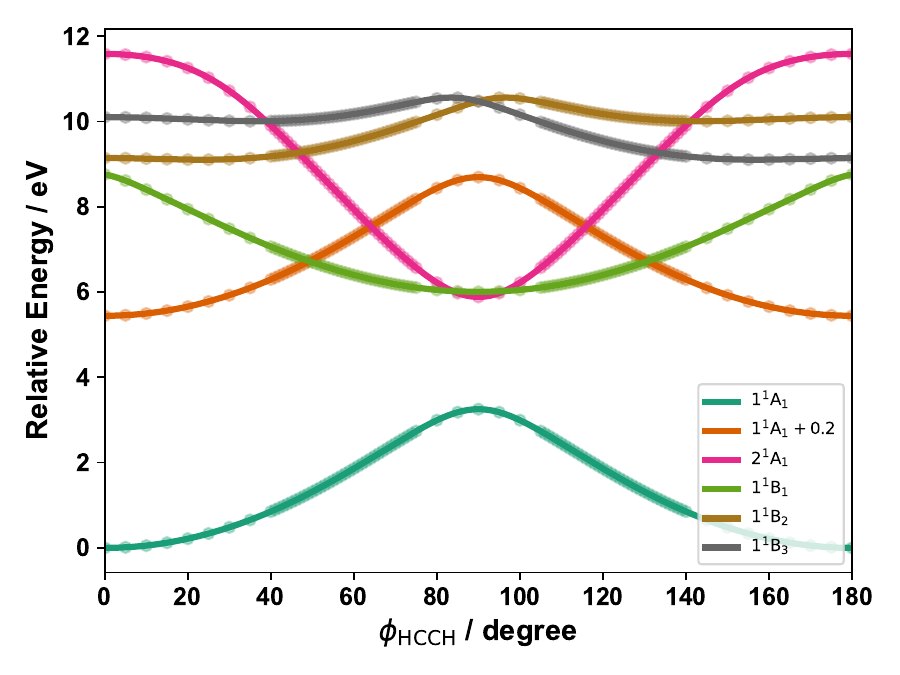}
	\caption{Potential energies curves (shifted by the ground state energy at $\phi_{\text{HCCH}}=0$)
of the five low-lying singlet states for \ce{C2H4} outside the optical cavity.
	Each state is labeled by the irrep of $\text{D}_{\text{2}}$ at  $\phi_{\text{HCCH}}=90^\circ$.
The symbol ``$^1A_1+0.2$'' refers to $|1\,^{1}\!\mathrm{A}_1\rangle \otimes |1_{\mathrm{P}}\rangle$ (or $|\text{S}_0\rangle \otimes |1_{\mathrm{P}}\rangle$)
obtained by shifting up $^1A_1$  by 0.2 au.
	}
	\label{fig:c2h4_cavity_free}
\end{figure}

\begin{figure}[!ht]
	\includegraphics[width=0.5\textwidth]{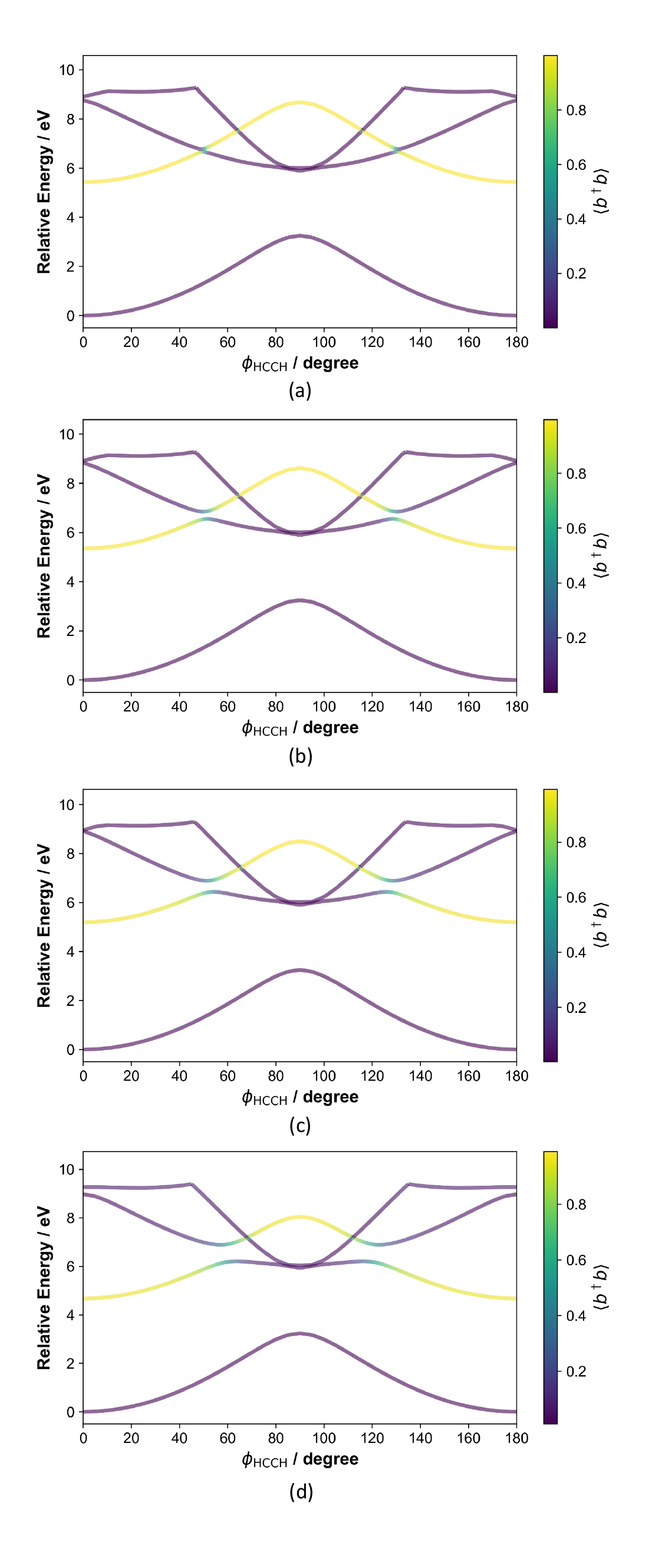}
	\caption{Potential energies curves (shifted by the ground state energy at $\phi_{\text{HCCH}}=0$) of the four low-lying singlet states for \ce{C2H4} inside an optical cavity with cavity parameter $\omega=0.2$ a.u. and coupling strength $|\lambda|=$ (a) 0.01, (b) 0.03, (c) 0.05 and (d) 0.10 a.u.
	The cavity mode is polarized along the \ce{C}-\ce{C} bond.
	The color indicates the average photon number $\langle b^\dagger b\rangle$ for each states.
	}
	\label{fig:c2h4}
\end{figure}

\begin{figure}[!ht]
	\includegraphics[width=0.9\textwidth]{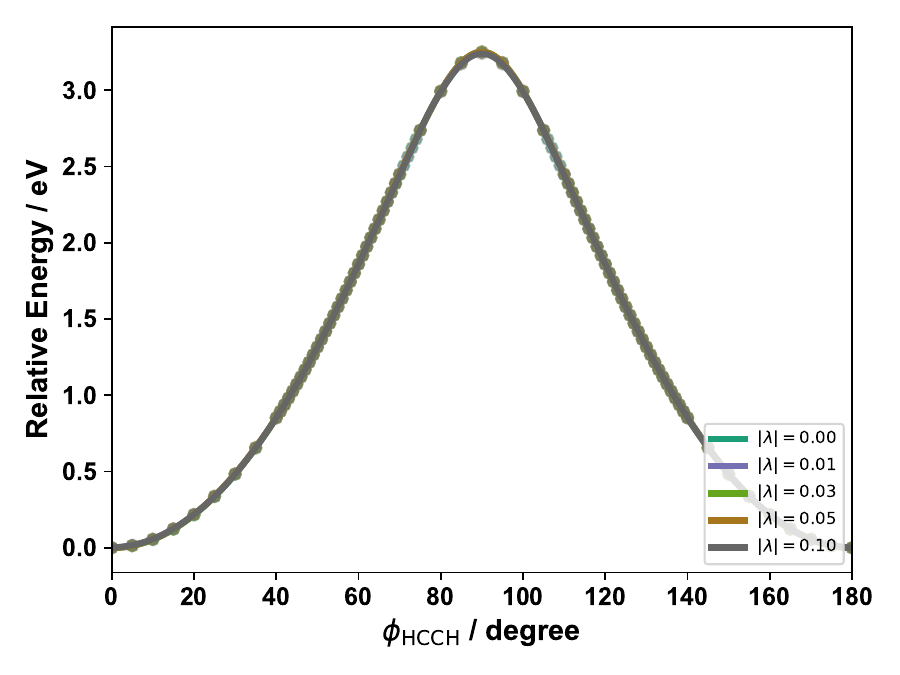}
	\caption{Potential energies curves (shifted by the ground state energy at $\phi_{\text{HCCH}}=0$) of the ground singlet state for \ce{C2H4} in an optical cavity with cavity parameter $\omega=0.2$ a.u. and various coupling strengths.
	The cavity mode is polarized along the \ce{C}-\ce{C} bond.
	}
	\label{fig:c2h4_gt}
\end{figure}

Figure~\ref{fig:c2h4_cavity_free} displays the five low-lying singlet states of \ce{C2H4}.
These states are identified as the five lowest in energy at the dihedral angle $\phi_{\text{HCCH}} = 90^\circ$ and are labeled by the irreducible representations of $D_{2}$.
The cavity frequency is set to $\omega = 0.2$~a.u.~(5.44~eV). With this choice but  in the absence of coupling ($|\bm{\lambda}| = 0.0$~a.u.), the polaritonic state  $|1\,^{1}\!\mathrm{A}_1\rangle \otimes |1_{\mathrm{P}}\rangle$ (or $|\text{S}_0\rangle \otimes |1_{\mathrm{P}}\rangle$) crosses 
the electronically excited states $|1\,^{1}\!\mathrm{B}_1\rangle \otimes |0_{\mathrm{P}}\rangle$ and $|2\,^{1}\!\mathrm{A}_1\rangle \otimes |0_{\mathrm{P}}\rangle$ 
at $\phi_{\text{HCCH}} \approx 49^\circ$ and $\phi_{\text{HCCH}} \approx 64^\circ$, respectively.
In the presence of cavity couplings, the three excited states will be mixed. The extent of mixture can be measured by the average PN 
shown in Fig.~\ref{fig:c2h4}. It can be seen that, although  at small $|\bm{\lambda}|$
the average PNs of the two participating polaritonic states 
deviate significantly from 0 or 1 only in a narrow region around each resonance,
 at $|\bm{\lambda}| = 0.10$~au the upper state formed near $\phi_{\text{HCCH}} = 40^\circ$
still carries an average PN well below $1.0$ when it approaches the third state. 
For comparison, the ground singlet state dominated by $|1\,^{1}\!\mathrm{A}_1\rangle \otimes |0_{\mathrm{P}}\rangle$
is hardly modified by the cavity, see Fg. \ref{fig:c2h4_gt}.

%
It can also be seen from Fig.~\ref{fig:c2h4} that  
the energy gap around the first crossing point ($\phi_{\text{HCCH}} \approx 49^\circ$)
increases considerably at larger coupling strengths,
whereas this is not the case around the second crossing point ($\phi_{\text{HCCH}} \approx 64^\circ$).
This can be understood from a symmetry argument: when the cavity photon is polarized along the \ce{C}-\ce{C} bond, 
the bilinear coupling $\hat{H}_{\mathrm{int}}$ transforms as the $\mathrm{B}_1$ irrep.
Consequently, the matrix element between $|1\,^{1}\!\mathrm{A}_1\rangle \otimes |1_{\mathrm{P}}\rangle$ and $|1\,^{1}\!\mathrm{B}_1\rangle \otimes |0_{\mathrm{P}}\rangle$ is non-vanishing, whereas the analogous coupling between $|1\,^{1}\!\mathrm{A}_1\rangle \otimes |1_{\mathrm{P}}\rangle$ and $|2\,^{1}\!\mathrm{A}_1\rangle \otimes |0_{\mathrm{P}}\rangle$ vanishes by symmetry.
Therefore, the states around $\phi_{\text{HCCH}}\approx 49^\circ$ exhibit an avoided crossing, 
while those around $\phi_{\text{HCCH}} \approx 64^\circ$ simply cross.
Furthermore, as $|\bm{\lambda}|$ increases, the stronger interaction arising from the first 
crossing pushes the second intersection towards larger angles, from approximately $64^\circ$ to $68^\circ$.
This reveals that when two resonance points lie close, an optical cavity has the possibility to 
significantly reshape the excited-state potential energy surfaces.
The consequences of this reshaping for chemical phenomena, including chemical reactivity, 
remain to be further explored.

\subsection{Barriers of proton transfer reactions } \label{sec:proton_transfer}
Proton transfer is ubiquitous in many fundamental chemical and biological processes.  
The ability to modulate and control its rate is indispensable for  
numerous quantum technological advances, such as the design of fuel cells, electrochemical devices, and solar energy harvesting.
In this subsection, we focus on the barriers of the proton transfer reactions in malonaldehyde and aminopropenal
when they are coupled to optical cavities.
All calculations in this subsection employ the cc-pVDZ basis set, with the same cavity settings 
($\omega = 3$~eV and $|\bm{\lambda}| = 0.1$~au) as used before\cite{CavityProtonTransferJACS,QED_LF_MP2}.
The geometries are taken from Ref.~\cite{CavityProtonTransferJACS}.
To reduce computational cost, 14 core electrons (the 1s orbitals of \ce{C}, \ce{N}, and \ce{O}, and the 2s orbitals of \ce{N} and \ce{O}) are frozen, leading to CAS(24e,88o).
The core-electron contribution is estimated via the energy difference between cavity-free 
CCSD (coupled-cluster with singles and doubles) calculations with and without the frozen electrons.

\begin{figure}[!ht]
	\includegraphics[width=0.9\textwidth]{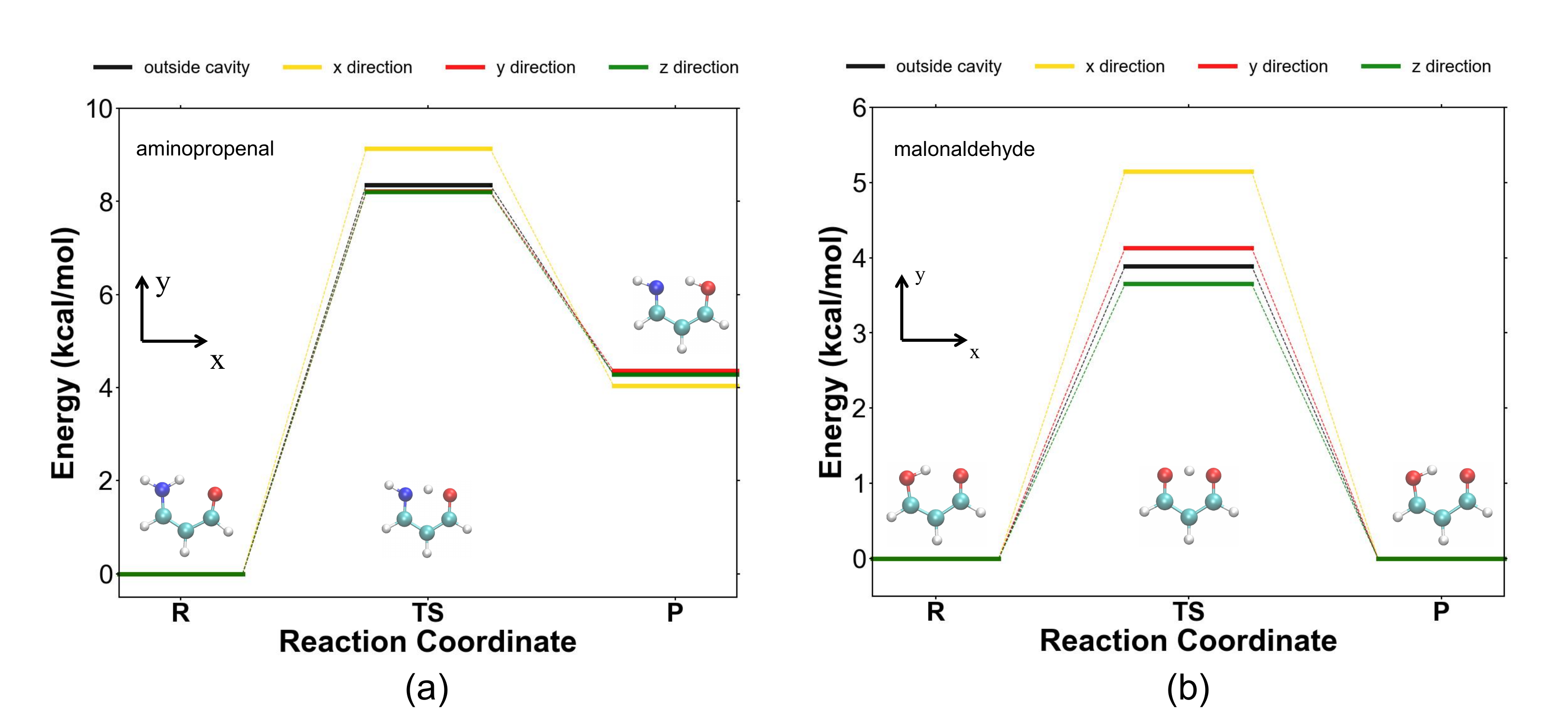}
	\caption{
		cQED-CS-iCIPT2 reaction diagrams for proton transfer in aminopropenal (panel a) and malonaldehyde (panel b) outside and inside the cavity.
		The calculations employ cavity parameters $\omega=3$ eV and $\lambda=0.1$ au, with the proton mode polarized along the $x$, $y$ and $z$ directions.
		Each panel contains the structures of the reactant (R), transition state (TS), and product (P) together with the coordinate frame.}
	\label{fig:cqed_proton_icipt2}
\end{figure}

The cQED-CS-iCIPT2 reaction diagrams for the proton transfers in malonaldehyde and aminopropenal are shown in 
Fig.~\ref{fig:cqed_proton_icipt2}. For malonaldehyde, 
when the cavity photon is polarized along the $x$ ($y$) axis, the reaction barrier increases by 1.26 (0.24) kcal/mol, 
a trend already apparent at the mean-field level (1.17 (0.15) kcal/mol),  see Table \ref{table:malon}. In contrast, 
the reaction barrier is reduced when the polarization is along the $z$ axis ($-0.24$ and $-0.07$~kcal/mol for cQED-CS‑iCIPT2 and CS‑HF, respectively).
The situation is different for aminopropenal. As can be seen from Table \ref{table:amino}, the reaction barrier is 
increased by 0.78~kcal/mol for the $x$-polarization but reduced by 0.14 and 0.16~kcal/mol for the $y$- and $z$-polarizations, respectively.
Moreover, the $x$-polarization  
stabilizes the product by 0.26~kcal/mol, whereas the $y$- and $z$-polarizations have only marginal effects ($-0.07$ and $0.01$~kcal/mol, respectively). 
These result indicate that proton transfer reactions (and many other reactions) can be fine-tuned by optical cavities. 

iCIPT2/cQED-CS‑iCIPT2 can be taken as reference to calibrate other methods. For instance, compared with iCIPT2, 
CCSD overestimates the reaction barrier of cavity-free  malonaldehyde by about 1.3~kcal/mol, whereas compared with 
cQED-CS‑iCIPT2, CCSD-22 overestimates the reaction barriers of cavitied malonaldehyde essentially by the same amounts (i.e., 
0.88, 1.10, and 1.30 kcal/mol for the $x$-, $y$-, and $z$-polarized photons, respectively; see Table.~\ref{table:malon}).
The same trend occurs also to the reaction barriers of aminopropenal, see Table \ref{table:amino}.
Close inspections reveal that over 99\% of the PN basis states determined automatically by cQED-CS‑iCIPT2 carry PNs 0 or 1, although 
the maximum PN reaches four. It can hence be concluded that 
the deviation of CCSD/CCSD-22 from iCIPT2/cQED-CS-iCIPT2 is primarily due to the inadequacy of CCSD in describing the 
electron-electron correlation.

\begin{table}[!ht]
	\centering
	\caption{
		Reaction energy barriers (in kcal/mol) for proton transfer in malonaldehyde outside and inside optical cavity calculated at different levels of theory.
			Extrapolation uncertainties of cQED-CS-iCIPT2 are given in parentheses.
	}
	\begin{threeparttable}
		\begin{tabular}{l|SSSS} \toprule
			Method
			& \multicolumn{1}{c}{outside cavity}
			& \multicolumn{1}{c}{x direction}
			& \multicolumn{1}{c}{y direction}
			& \multicolumn{1}{c}{z direction}  \\ \toprule
			cQED-HF                  & 9.27  & 10.44  & 9.41 & 9.19 \\
			cQED-CCSD-22\tnote{a}    & 5.19  & 6.03  & 5.23  & 4.95   \\ 
			cQED-CS-iCIPT2  & 3.89(13)  & 5.15(10)  & 4.13(5)  & 3.65(7)  \\ 
			\bottomrule
		\end{tabular}
		\begin{tablenotes}
			\item[a] Ref.~[\citenum{CavityProtonTransferJACS}].
			The cluster operator is restricted to single and double electronic excitations together with their interaction with two photons.
		\end{tablenotes}
	\end{threeparttable} \label{table:malon}
\end{table}

\begin{table}[!ht]
	\tiny
	\centering
	\caption{
		Reaction energy barriers (TS in kcal/mol) and reaction energies ($\Delta E$ in kcal/mol) for proton transfer in aminopropenal outside and inside cavity calculated at different levels of theory.
	}
	\begin{threeparttable}
		\begin{tabular}{l|SS|SS|SS|SS} \toprule
			\multirow{2}{*}{Method}
			& \multicolumn{2}{c|}{outside cavity}
			& \multicolumn{2}{c|}{x direction}
			& \multicolumn{2}{c|}{y direction}
			& \multicolumn{2}{c}{z direction}   \\ \cline{2-9}
			& \multicolumn{1}{c}{TS}
			& \multicolumn{1}{c|}{$\Delta E$}
			& \multicolumn{1}{c}{TS}
			& \multicolumn{1}{c|}{$\Delta E$}
			& \multicolumn{1}{c}{TS}
			& \multicolumn{1}{c|}{$\Delta E$}
			& \multicolumn{1}{c}{TS}
			& \multicolumn{1}{c}{$\Delta E$} \\  \toprule
			CS-HF             & 16.64  & 8.62  & 18.64  &  9.56 & 17.22 & 8.99 & 16.63  & 8.69   \\
			cQED-CCSD-22\tnote{a}   & 9.65  & 4.54   & 10.72 & 4.67  & 9.71  & 4.32  & 9.32  & 4.34   \\ 
			cQED-CS-iCIPT2 & 8.36(14)  & 4.29(19)   & 9.14(7)  & 4.04(10)  & 8.22(4)  & 4.36(5)  & 8.20(8)  & 4.28(13)   \\ 
			\bottomrule
		\end{tabular}
		\begin{tablenotes}
			\item[a] Ref.~[\citenum{CavityProtonTransferJACS}].
			The cluster operator is restricted to single and double electronic excitations together with their interaction with two photons.
		\end{tablenotes}
	\end{threeparttable} \label{table:amino}
\end{table}

\newpage
\clearpage

\subsection{Low-Lying States of Polyacenes} \label{sec:Polyacenes}

As a final example, we study the low-lying polaritonic states of $k$-polyacenes ($k \in [2, 10]$).
The active space comprises the $2p_z$ orbitals of the carbon atoms, corresponding up to a CAS(42e,42o).
Cavity-free iCISCF\cite{iCISCF} orbitals for the ground singlet states of $k$-polyacenes  are used in cQED-CS-iCIPT2/CAS(42e,42o) calculations.
The cavity photon frequencies are set to the $\text{S}_0 \rightarrow \text{S}_1$ electronic excitation energies (see the Supporting Information).

\begin{figure}[!ht]
	\includegraphics[width=1.0\textwidth]{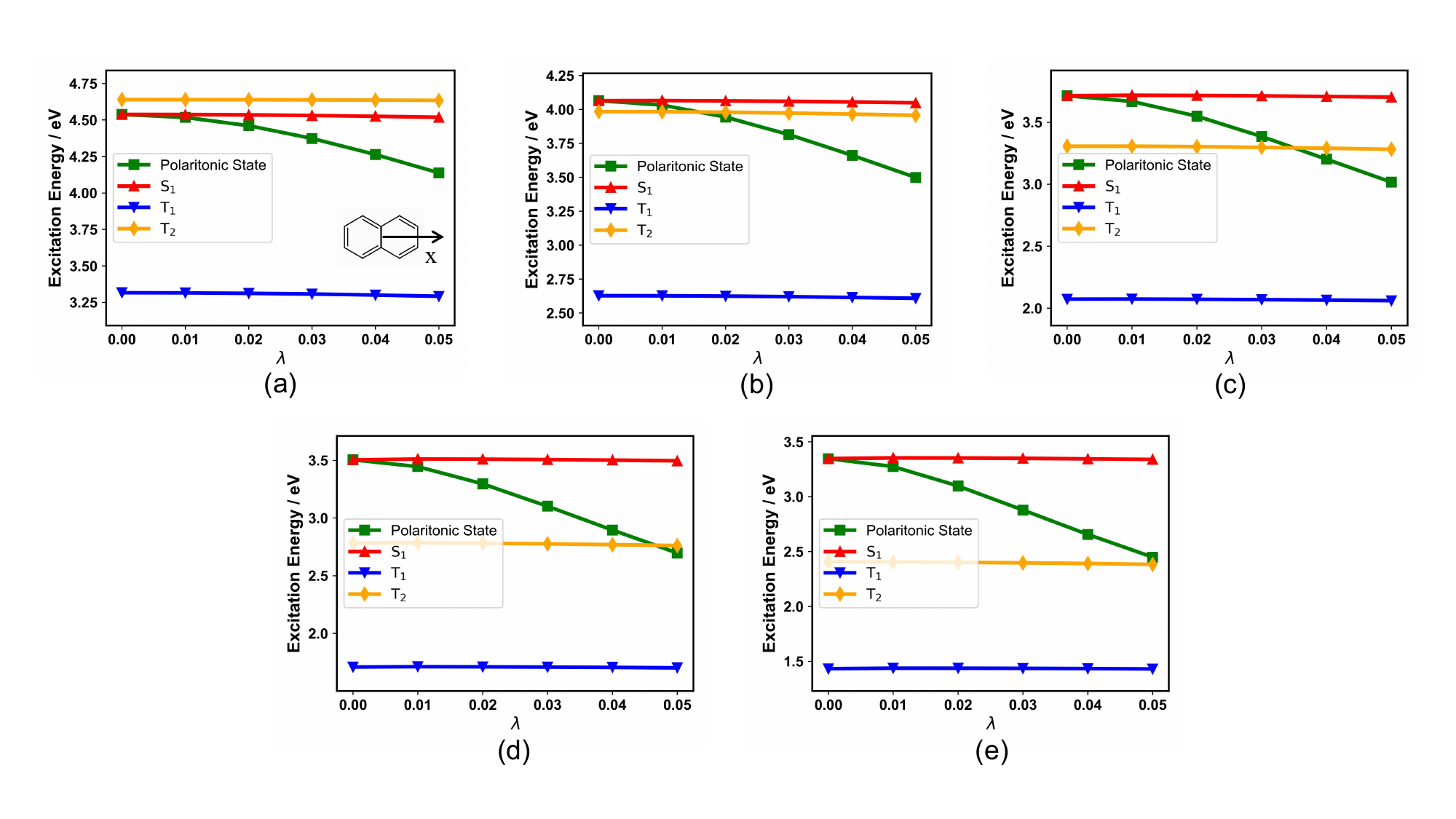}
	\caption{Excitation energies (in eV) of the low-lying states of $k$-polyacenes at various coupling strengths $|\lambda|$. 
 Subpanels (a)--(e) refer to $k=2$ to 6, respectively. The coupling vector is oriented along the long axis of the molecule.}
	\label{fig:polyacene}
\end{figure}

\begin{figure}[!ht]
	\includegraphics[width=1.0\textwidth]{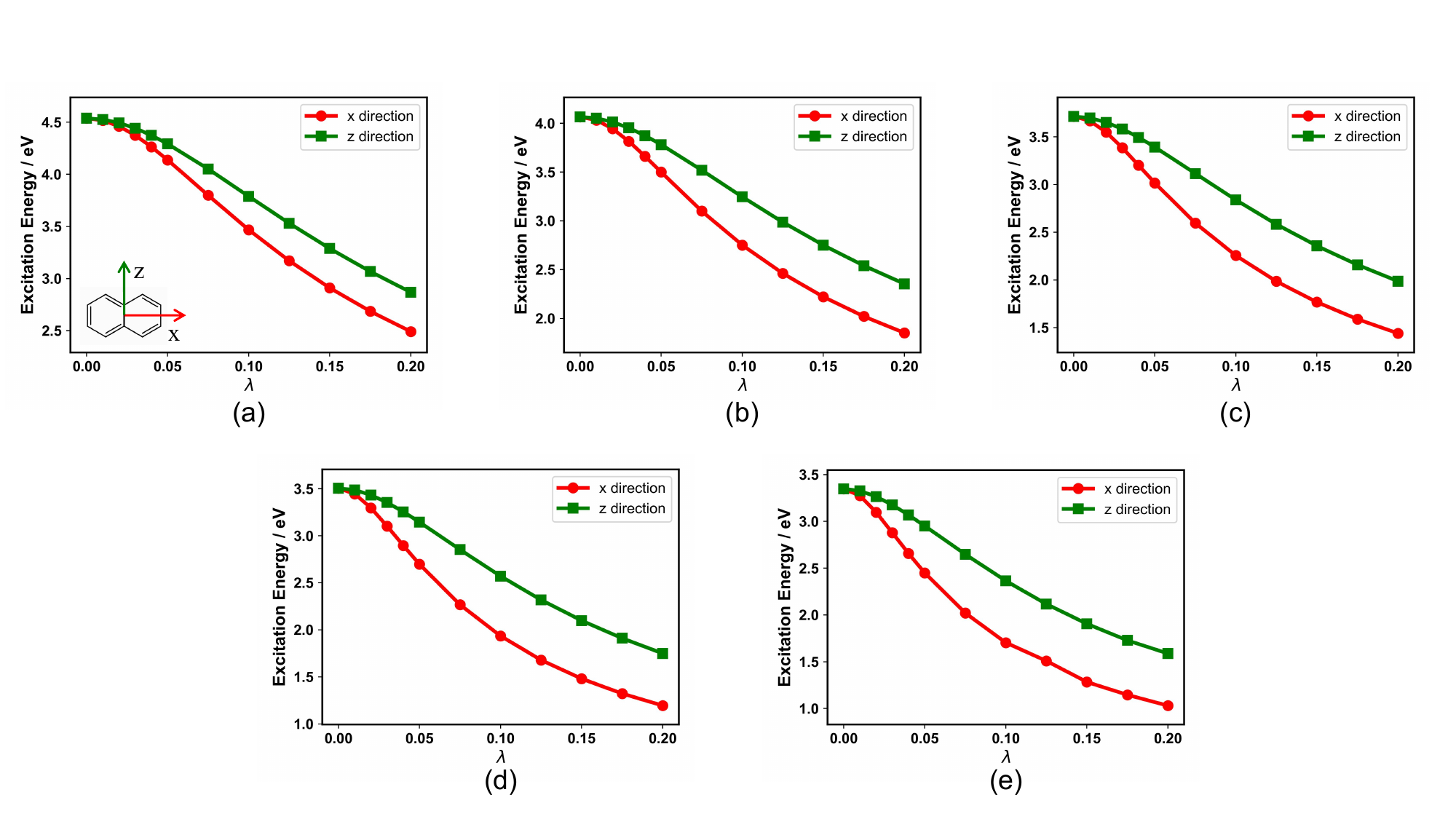}
	\caption{Excitation energies (in eV) of the polaritonic states of $k$-polyacenes at various coupling strengths $|\lambda|$.
     Subpanels (a)--(e) refer to $k=2$ to 6, respectively. Two distinct orientations of the coupling vector are tested.}
	\label{fig:polyacene_polar}
\end{figure}

\begin{figure}[!ht]
	\includegraphics[width=1.0\textwidth]{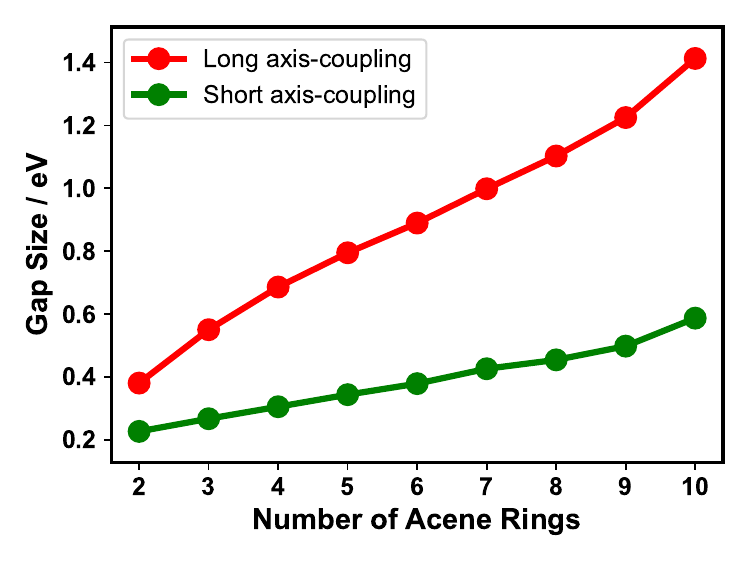}
	\caption{Energy separations (in eV) between the $S_1$ and  polaritonic states of $k$-polyacenes at coupling strength $|\lambda|=0.05$ au
but with different orientations of the coupling vector.}
	\label{fig:polyacene_gap}
\end{figure}

The electronic and polaritonic excitation energies of $k$-polyacenes ($k\in[2,6]$) are plotted in Fig.~\ref{fig:polyacene} as function of the coupling strength.
It can be seen that all states except the polaritonic state are affected almost equally by the cavity photon up to a coupling strength of $0.05$ au.
This feature no longer holds for larger coupling strengths, as shown in the Supporting Information.
Furthermore, the excitation energy of the polaritonic state decreases with increasing coupling strength, which may lead to state crossings.
However, because the $\text{S}_1$--$\text{T}_2$ gap increases with the number of acene rings, 
state inversion occurs at larger coupling strengths (see Fig.~\ref{fig:polyacene}), demonstrating the influence of the molecular structure on the polaritonic behavior.
The excitation energies of the polaritonic states also depend on the orientation of the coupling vector.
As shown in Fig.~\ref{fig:polyacene_polar}, the polaritonic states exhibit 
lower excitation energies for the polarization direction along the long molecular axis than for that along the short molecular axis.
The splitting (the energy difference between $\text{S}_1$ and the polaritonic state) 
exhibits a remarkable linear increase with the number of rings at  
a coupling strength of $0.05$, as shown in Fig.~\ref{fig:polyacene_gap}, 
again highlighting the influence of the molecular structure on polaritonic behavior.
Notably, the present cQED-CS-iCIPT2 has accessed larger polyacenes than cQED-DMRG\cite{QED_DMRG} 
(which stops at $k=5$ but with a similar linear trend.


\newpage
\clearpage

\section{Conclusions}\label{sec:conclusion}

We have extended the near-exact iCIPT2 method to ground and excited states of strongly coupled light–matter systems, 
resulting in the cQED-PN/CS-iCIPT2 approaches.
Leveraging the modular design and template metaprogramming techniques of the MetaWave platform, the 
implementation is achieved by first introducing a graded configuration space and 
then adapting each key step of iCIPT2 into intra- and inter-subspace components, so as to make maximal reuse of 
the existing code. A distinctive capability of cQED-PN/CS-iCIPT2 is its automatic convergence 
of the photon-number basis expansion for electron-photon correlation, achieved concurrently with the selection of important electronic configurations
for electron-electron correlation. 
This is numerically validated across a diverse set of challenging systems, 
including \ce{N2} bond dissociation, \ce{C2H4} torsion, proton transfer in malonaldehyde and aminopropenal, 
and low-lying excited states of polyacenes under strong cavity coupling. 
Although the coupling strengths employed here are beyond current experimental reach, 
they serve to unambiguously establish the robustness and efficiency of the cQED-PN/CS-iCIPT2 in regimes 
where conventional approaches fail. The MPI parallelization of cQED-PN/CS-iCIPT2, 
implemented via the unified framework already available in MetaWave\cite{MPI_iCIPT2}, will 
substantially extend their applicability to chemically and photochemically relevant systems 
that currently remain accessible only to approximate methods.

\section*{Acknowledgments}
This work was supported by National Natural Science Foundation of China  (Grant Nos. 22503051 and 22373057) and by Shandong Postdoctoral Science Foundation.

\section*{Supporting Information}
The Supporting Information contains discussions on different treatment of the DSE term and the cavity photon frequencies used in polyacenes.

\section*{Data and Software Availability}
The original data, testing and analyzing scripts, typical input and output file have been made publicly available on the GitHub repository SI\_cQEDiCIPT2\cite{SICQEDICIPT2}.
The software (MetaWave) is closed-source.
Access to the MetaWave package may be granted upon reasonable request to the corresponding authors.

\section*{Conflicts of interest}
There are no conflicts to declare.




\newpage
\clearpage

\bibliography{iCI}

\end{document}